\documentclass[fleqn,usenatbib]{mnras}

\usepackage{newtxtext,newtxmath}

\usepackage[T1]{fontenc}
\usepackage{ae,aecompl}
\usepackage{natbib}

\def\fesc{\ifmmode f_{\rm esc} \else $f_{\rm esc}$\fi}

\usepackage{graphicx}	

\title[CNO abundances in LyC leaking galaxies]{Abundances of CNO elements in
$z \sim $ 0.3 -- 0.4 LyC leaking galaxies}

\author[Y. I. Izotov et al.]{
Y. I. Izotov$^{1}$\thanks{Corresponding author: yizotov@bitp.kiev.ua},
D. Schaerer$^{2,3}$, 
G. Worseck$^{4}$, 
D. Berg$^{5}$,
J. Chisholm$^{5}$, 
\newauthor 
~S. Ravindranath$^{6}$, T. X. Thuan$^{7}$
\\
$^{1}$Bogolyubov Institute for Theoretical Physics,
National Academy of Sciences of Ukraine, 14-b Metrolohichna str., Kyiv,
03143, Ukraine,\\
E-mail: yizotov@bitp.kiev.ua\\
$^{2}$Observatoire de Gen\`eve, Universit\'e de Gen\`eve, 
51 Ch. des Maillettes, 1290, Versoix, Switzerland,\\
E-mail: daniel.schaerer@unige.ch\\ 
$^{3}$IRAP/CNRS, 14, Av. E. Belin, 31400 Toulouse, France\\
$^{4}$ Institut f\"ur Physik und Astronomie, Universit\"at Potsdam,
Karl-Liebknecht-Str. 24/25, D-14476 Potsdam, Germany,\\
E-mail: gworseck@uni-potsdam.de\\
$^{5}$Astronomy Department, University of Texas at Austin,
2515 Speedway, Stop C1400 Austin, TX 78712-1205, USA, \\
E-mail: daberg@austin.utexas.edu, chisholm@austin.utexas.edu\\
$^{6}$Space Telescope Science Institute, 3700 San Martin Drive, Baltimore, MD,
21218, USA, \\
E-mail: swara@stsci.edu\\
$^{7}$Astronomy Department, University of Virginia, P.O. Box 400325, 
Charlottesville, VA 22904-4325, USA. \\
E-mail: txt@virginia.edu\\
}

\date{Accepted XXX. Received YYY; in original form ZZZ}

\pubyear{2022}

\begin{document}
\label{firstpage}
\pagerange{\pageref{firstpage}--\pageref{lastpage}}
\maketitle

\begin{abstract}
We present observations with the Space Telescope Imaging Spectrograph (STIS)
onboard the {\sl Hubble Space Telescope} of eleven Lyman continuum (LyC) leaking
galaxies at redshifts, $z$, in the range 0.29\,--\,0.43, with oxygen abundances
12\,+\,log(O/H)\,=\,7.64\,--\,8.16, stellar masses 
$M_\star$\,$\sim$\,10$^{7.8}$\,--\,10$^{9.8}$\,M$_\odot$ and
O$_{32}$\,=\,[O~{\sc iii}]\,$\lambda$5007/[O~{\sc ii}]\,$\lambda$3727 of
$\sim$\,5\,--\,20 aiming to detect the C~{\sc iii}]\,$\lambda$1908 emission
line. We combine these observations with the optical
Sloan Digital Sky Survey (SDSS) spectra for the determination of
carbon, nitrogen and oxygen abundances.
Our sample was supplemented by thirty one galaxies from the literature,
for which carbon, nitrogen and oxygen abundances can be derived from the
{\sl HST} and SDSS spectra. These additional galaxies, however,
do not have LyC observations.
We find that log(C/O) for the entire sample at
12\,+\,log(O/H)~$<$~8.1 does not depend on metallicity,
with a small dispersion of $\sim$\,0.13 dex around the average value
of $\sim$\,$-$0.75 dex. On the other hand, the log(N/O) in
galaxies at $z$\,$>$\,0.1, including LyC leakers, is systematically higher
compared to the rest of the sample with lower metallicity.
We find that log(C/O) slightly decreases
with increasing $M_\star$ from $\sim$\,$-$0.65 at $M_\star$\,=\,10$^6$\,M$_\odot$
to $\sim$\,$-$0.80 at $M_\star$\,=\,10$^9$\,--\,10$^{10}$\,M$_\odot$, whereas
log(N/O) is
considerably enhanced at $M_\star$\,$>$\,10$^8$\,M$_\odot$. The origin of these
trends remains basically unknown. A possible solution would be to assume
that the upper mass limit of the stellar initial mass function (IMF) in more
massive galaxies is higher. This would result in
a higher production of oxygen and a larger fraction of massive stars with stellar
wind polluting the interstellar medium with nitrogen.
\end{abstract}




\section{Introduction}\label{intro}

The most abundant heavy element species, carbon, nitrogen and oxygen, play
a crucial role in describing the chemical enrichment of galaxies.
This is because oxygen is produced only by short-lived massive stars
with $M_\star$ $>$ 10 M$_\odot$, whereas
carbon and nitrogen are synthesized by both massive stars and longer-lived
stars of lower masses, resulting in variations of C/O and N/O abundance ratios
with time. Since both the oxygen and carbon are primary elements, their
abundance ratio variations as a function of oxygen abundance are expected
to be flat.
As for nitrogen, it can be produced in both primary and secondary processes,
but the contribution of secondary nitrogen at low metallicities is small and
no significant dependence of the N/O abundance ratio on oxygen abundance is
expected. Indeed, it was found in many studies that the N/O abundance ratio at
low oxygen abundances 12\,+\,log(O/H) $\leq$ 8.0 is constant, suggesting that the
N production is primary \citep[e.g. ][]{I06}, with however a significant scatter
($\pm$ 0.5 dex).
This scatter may be due to the time-dependent nature of N/O abundance ratios
and/or different star formation efficiencies setting different fundamental
plateau values \citep*[e.g. ][]{H00,Be20}. Furthermore, \citet{I06} suggested
that the scatter can also be due to local nitrogen pollution by Wolf-Rayet stars.
At higher metallicities, the secondary
production of nitrogen becomes important and thus the N/O abundance ratio
increases with increasing oxygen abundance.

As for carbon, abundances of its ions can be derived from the emission of weak
recombination lines in the optical range, but this is practically possible only
in bright H~{\sc ii} regions, with high oxygen abundances 12\,+\,log(O/H) $\ga$ 8.0
\citep[e.g. ][]{Es14}.
At lower metallicities only collisionally-excited forbidden lines in the UV
range can be used to derive the carbon abundance. However, rest-frame UV
observations of the galaxies in the local Universe are only possible from space.

Several fundamental studies on the determination of the carbon abundances
have been made with the {\sl Hubble Space Telescope} ({\sl HST}) starting from
the pioneering work by \citet{Ga95}, who found that C/O abundance ratio
in low-metallicity dwarf star-forming galaxies (SFGs) linearly increases with
oxygen abundance O/H. However, this finding has not been confirmed in later
studies of low-metallicity SFGs. It is also inconsistent with the constant
C/O abundance ratio in Milky Way halo stars with metallicities in the same
range \citep[e.g. ][]{Ak04,Fa09}.
Thus, \citet{IT99} re-analysed observations of five galaxies from
the \citet{Ga95} sample and added one more galaxy from \citet{Th99}, all
with 12\,+\,log(O/H) $\leq$ 7.6, and found log(C/O) and log(C/N) to be constant, at
$-$0.78\,$\pm$\,0.03 and 0.82\,$\pm$\,0.04, respectively. These constant abundance
ratios imply that C and N at low metallicities have a primary origin.

  \begin{table*}
  \caption{Coordinates and characteristics derived from the COS and SDSS spectra
\label{tab1}}
\begin{tabular}{lrrcrcccc} \hline
Name&R.A.(2000.0)&Dec.(2000.0)&$z$$^{\rm a}$&O$_{32}$$^{\rm b}$&log($M_\star$/M$_\odot$)$^{\rm c}$&12+log(O/H)$^{\rm d}$&log(N/O)&$f_{\rm esc}$(LyC) \\ \hline
J0901$+$2119&09:01:45.61&$+$21:19:27.70&0.29932&20.0& 9.80&8.16$\pm$0.02&$-$1.22$\pm$0.07&0.027$\pm$0.007\\
J0925$+$1403&09:25:32.37&$+$14:03:13.06&0.30121& 5.9& 8.91&7.91$\pm$0.01&$-$1.12$\pm$0.04&0.078$\pm$0.008\\
J1011$+$1947&10:11:38.28&$+$19:47:20.90&0.33219&19.2& 9.00&7.99$\pm$0.01&$-$0.85$\pm$0.09&0.114$\pm$0.018\\
J1152$+$3400&11:52:04.88&$+$34:00:49.88&0.34197& 3.9& 9.59&8.00$\pm$0.02&$-$1.17$\pm$0.10&0.132$\pm$0.011\\
J1154$+$2443&11:54:48.85&$+$24:43:33.02&0.36900&11.5& 8.20&7.65$\pm$0.02&$-$1.30$\pm$0.25&0.460$\pm$0.020\\
J1243$+$4646&12:43:00.63&$+$46:46:50.40&0.43166&11.2& 7.80&7.89$\pm$0.01&$-$1.14$\pm$0.16&0.726$\pm$0.097\\
J1248$+$4259&12:48:10.48&$+$42:59:53.60&0.36300&19.7& 8.20&7.64$\pm$0.01&$-$1.07$\pm$0.12&0.022$\pm$0.007\\
J1256$+$4509&12:56:44.15&$+$45:09:17.00&0.35311&16.3& 8.80&7.87$\pm$0.01&$-$1.01$\pm$0.12&0.380$\pm$0.057\\
J1333$+$6246&13:33:03.96&$+$62:46:03.78&0.31816& 4.8& 8.50&7.76$\pm$0.02&$-$1.54$\pm$0.19&0.056$\pm$0.015\\
J1442$-$0209&14:42:31.39&$-$02:09:52.03&0.29364& 7.9& 8.96&7.93$\pm$0.01&$-$1.25$\pm$0.09&0.074$\pm$0.010\\
J1503$+$3644&15:03:42.83&$+$36:44:50.75&0.35562& 4.9& 8.22&7.95$\pm$0.01&$-$1.25$\pm$0.09&0.058$\pm$0.006\\
\hline
\end{tabular}

\hbox{$^{\rm b}$$z$ is the redshift.}

\hbox{$^{\rm b}$O$_{32}$ is the extinction-corrected ratio $I$([O~{\sc iii}]$\lambda$5007)/$I$([O~{\sc ii}]$\lambda$3727).}

\hbox{$^{\rm c}$$M_\star$ is the stellar mass.}

\hbox{$^{\rm d}$Oxygen abundance derived by the direct $T_{\rm e}$ method.}

  \end{table*}

  \begin{table}
  \caption{{\sl HST}/STIS observations \label{tab2}}
  \begin{tabular}{lccr} \hline
\multicolumn{1}{c}{Name}&\multicolumn{1}{c}{Date}&\multicolumn{2}{c}{Exposure time (s)} \\ 
    &    &Acquisition&Science \\ \hline
J0901$+$2119&2021-04-13&440     & 12539 \\
J0925$+$1403&2021-03-21&200     &  7387 \\
J1011$+$1947&2021-04-17&340     & 12734 \\
J1152$+$3400&2021-02-04&140     &  7600 \\
J1154$+$2443&2018-04-28&360     & 10418 \\
            &2019-01-02&360     &  9290 \\
J1243$+$4646&2021-01-30&420     & 13219 \\
J1248$+$4259&2021-02-03&300     & 10339 \\
J1256$+$4509&2022-05-25&580     &  7012 \\
            &2022-06-28&290     &  4376 \\
J1333$+$6246&2021-12-22&180     & 11271 \\
J1442$-$0209&2020-06-18&180     &  7406 \\
J1503$+$3644&2022-02-05&180     &  7500 \\
\hline
\end{tabular}



  \end{table}

\begin{figure*}
\includegraphics[angle=0,width=0.99\linewidth]{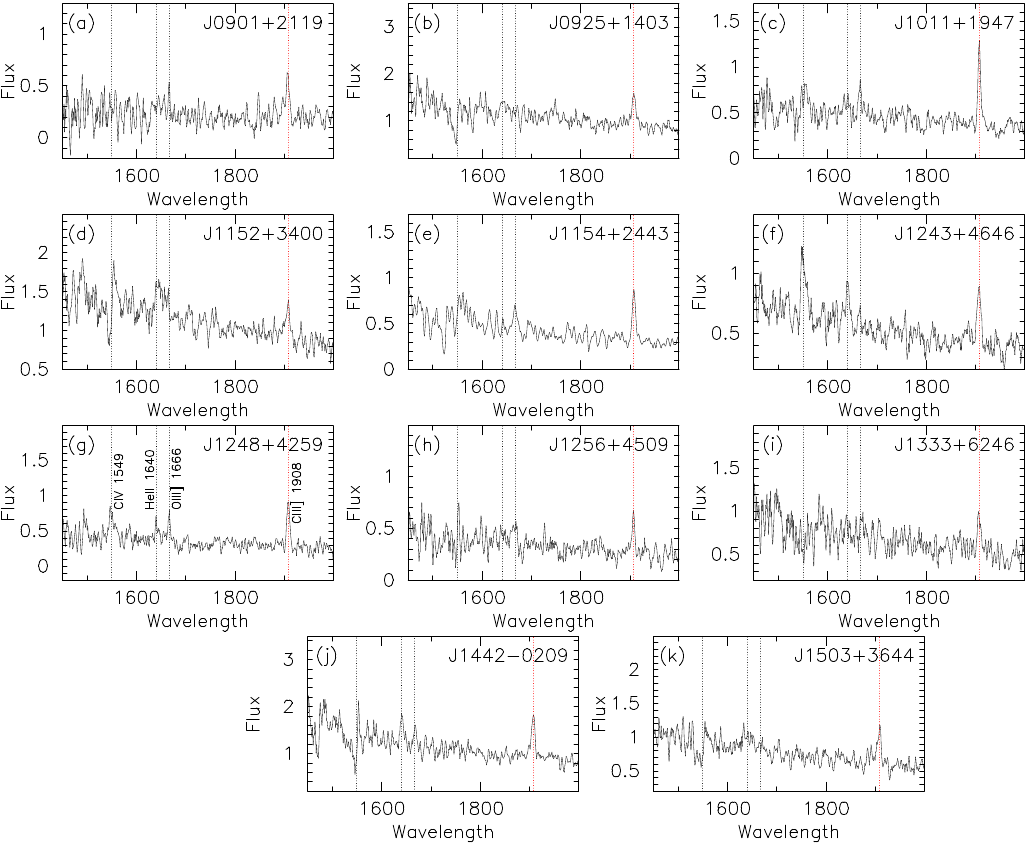}

\caption{{\sl HST}/STIS spectra of LyC leaking galaxies with the G230L grating.
  The spectra are shifted to rest-frame wavelengths.
The strongest lines are labelled in {\bf (g)} and their locations
are marked by vertical red dotted lines (C~{\sc iii}] $\lambda$1908) and by
vertical black dotted lines (other lines). Fluxes and wavelengths are
expressed in 10$^{-16}$ erg s$^{-1}$ cm$^{-2}$ \AA$^{-1}$ and \AA, respectively.
\label{fig1}}
\end{figure*}

All of the early {\sl HST} observations were obtained with the Faint Object
Spectrograph (FOS). Subsequent observations with the Cosmic Origins Spectrograph
(COS) and Space Telescope Imaging Spectrograph (STIS) onboard {\sl HST}
considerably increased the number of galaxies with a determined C/O abundance
ratio.

Using COS observations of 21 dwarf star-forming galaxies with $z$ $<$ 0.1,
\citet{Be16,Be19} 
found no trend in C/O at low metallicity and a constant C/N ratio but with
significant scatter over a larger range in oxygen abundance, indicating that
carbon is predominantly produced by the same nucleosynthetic mechanisms as
nitrogen. \citet{Be19} derived log(C/O)\,=\,--0.71 and log(C/N)\,=\,0.75 with
dispersion $\sigma$\,=\,0.17 and 0.25, respectively. They also found a declining
C/O ratio with increasing baryonic mass of the galaxy, due to increasing
effective oxygen yields.

\citet{Se17} obtained {\sl HST}/COS spectra of 10 nearby galaxies at $z$ $<$ 1
with active ongoing star formation to study the variations in
C~{\sc iii}] $\lambda$1907, 1909\AA\ equivalent widths (hereafter
C~{\sc iii}] $\lambda$1908\AA). Combining with archival local
samples, they found that C~{\sc iii}]
emission lines with equivalent widths $>$~5\AA\ are observed only in galaxies
with 12 + log(O/H) $\la$ 8.4. On average, their EW(C~{\sc iii}]) values are
similar to those obtained by \citet{Be16,Be19} and in high-$z$ star-forming galaxies.
\citet{Se17} derived the C/O abundance ratios in their galaxies, but no analysis
of these values and a comparison with other studies have been done.

\citet{Ra20} analyzed the {\sl HST}/STIS near-UV (NUV) spectra for a sample
of 10 ``green pea'' (GP) galaxies at higher redshifts, 0.1~$\leq z \leq$~0.3.
They found that the C~{\sc iii}] EWs of GPs are in the range 2~--~10\AA,
overlapping with the range of values seen in low-$z$ and $z$~$>$~2 SFGs, but
not reaching the high EW(C~{\sc iii}]) values ($>$ 15\AA) seen in some
$z$ $>$ 2 SFGs \citep[e.g. ][]{Ll22} and in many SFGs by \citet{Be16,Be19}.
\citet{Ra20} also determined the C/O abundance ratios in their galaxies but
did not analyze or compare their values to other data sets.

In all low-$z$ studies of the carbon abundance considered above, the redshifts
of the galaxies are too low for {\sl HST} to observe the restframe wavelength
range of the Lyman Continuum (LyC).
On the other hand, it is important to study the
CNO abundances in the LyC leaking galaxies at higher redshift
to find common properties and possible differences associated with the
 LyC leakage with non-leaking galaxies and the lower-redshift galaxies,
for which observations of the LyC are not possible. The first study
of the carbon abundance in a LyC leaker
was done for J1154$+$2443 with a very high escape fraction
$f_{\rm esc}$(LyC) = 46 per cent, by \citet{Sc18}, who obtained
log(C/O)\,=\,--0.84 $\pm$ 0.06 from the C~{\sc iii}]\,$\lambda$1908 and
[O~{\sc iii}]\,$\lambda$4959,5007 line ratio. To increase the statistics
of LyC leaking galaxies with a derived carbon abundance, we consider in
this paper {\sl HST}/STIS observations of a sample of 10 LyC leakers from
\citet{I16a,I16b,I18a,I18b} with $f_{\rm esc}$(LyC) in the range 2\,--\,72 per cent
and very compact morphology. A main goal of these observations
was to study how high-ionization emission lines He~{\sc ii}\,$\lambda$1640 and
C~{\sc iv}\,$\lambda$1549, 1551 are linked to LyC leakage, as done by \citet{Sc22}.
But they allow also to derive carbon abundances in galaxies with higher stellar
masses, compared to that of low-$z$
galaxies studied by \citet{Be16,Be19}, and in a wide range of the LyC
escape fraction. That is the aim of the present study. All these galaxies were initially selected from the
Sloan Digital Sky Survey (SDSS). Their coordinates, redshifts,
O$_{32}$\,=\,[O~{\sc iii}]\,$\lambda$5007/[O~{\sc ii}]\,$\lambda$3727 flux
ratios, stellar masses,
and oxygen abundances are given in Table~\ref{tab1}. We also included the
galaxy J1154$+$2443 from the \citet{I16a,I16b,I18a,I18b} sample, which was
observed in program GO~15433 (PI: D.\ Schaerer) with the same STIS setup and
reduced in a similar manner \citep{Sc18}. However, \citet{Sc18} derived a
C/O abundance ratio from only a four-orbits exposure, whereas
we use the final spectrum of J1154$+$2443 with twice the exposure time.

The {\sl HST}/STIS observations of the LyC leaking galaxies are discussed in
Section~\ref{sec:obs}. The method for the determination of element abundances
is described in Section~\ref{sec:method}. In Section~\ref{sec:abund} we discuss
the derived CNO abundances and compare them with the results of previous
studies. We summarize our findings in Section~\ref{sec:summary}.

\section{{\sl HST}/STIS observations and data 
reduction}\label{sec:obs}

{\sl HST}/STIS spectroscopy of the 10 LyC leakers was obtained
in program GO~15941 (PI: D.\ Schaerer) during the period June 2020 --
June 2022 \citep{Sc22}. The observational details are presented in
Table \ref{tab2}. We also
included in the Table the galaxy J1154$+$2443, for which some results of STIS
spectroscopy were published in \citet{Sc18}.

The galaxies were directly acquired by STIS/CCD optical imaging using an image
obtained with the long-pass filter F28X50LP. The spectra were obtained with the
STIS/NUV-MAMA
low-resolution grating G230L, the central wavelength 2376\AA, and the
52~arcsec~$\times$~0.5~arcsec slit, resulting in a spectral resolution $R$ = 750
and wavelength coverage of 1570 - 3180\AA.
This setup allows to gather almost all light from our compact galaxies
\citep{I16a,I16b,I18a,I18b}, which have small angular sizes, $\sim$1 arcsec
in diameter, according to SDSS photometry.
The individual sub-exposures have somewhat different (by less than 1\AA)
spectral ranges and they were
reduced to a common wavelength range, which is the same for all exposures,
co-added using the {\sc iraf} {\it dispcor} routine, and the one-dimensional
(1D) spectrum was extracted using {\sc iraf}'s {\it apall} routine.

The rest-frame UV spectra of all galaxies are shown in Fig.~\ref{fig1}. The
locations of the four brightest lines in all panels are identified by vertical
black (highlighting the C~{\sc iv}~$\lambda$1550, He~{\sc ii}~$\lambda$1640, and
O~{\sc iii}]~$\lambda$1666\AA) and red dotted lines (highlighting the
C~{\sc iii}] lines) and are labelled in (g). The lines in some spectra
are weak and barely detected or non-detected. The only line which is clearly
detected in all spectra, is the C~{\sc iii}]~$\lambda$1908\AA\ emission line,
which, in fact, is a blended doublet. This allows us to
derive carbon abundance in all studied galaxies using {\sl HST} UV and SDSS
optical spectra.

The O~{\sc iii}]~$\lambda$1666\AA\ and C~{\sc iii}]~$\lambda$1908\AA\ emission
line fluxes, $F$(O~{\sc iii}]~$\lambda$1666\AA) and
$F$(C~{\sc iii}]~$\lambda$1908\AA), and
rest-frame equivalent widths EW(O~{\sc iii}]~$\lambda$1666\AA) and
EW(C~{\sc iii}]~$\lambda$1908\AA) have been measured using the {\sc iraf}
{\it splot} routine. The flux errors were obtained using eq.~1 by \citet{Be19}.
The results of measurements are shown in Table~\ref{tab3} together with the
observed fluxes $F$(H$\beta$) and rest-frame equivalent widths EW(H$\beta$)
of the H$\beta$ emission line, which are taken from \citet{I16a,I16b,I18a,I18b}.
We note that the equivalent widths EW(C~{\sc iii}]~$\lambda$1908\AA) of
$z$ $\sim$ 0.3 -- 0.4 LyC leakers are high. They are similar to those in
the lower-redshift galaxies studied by \citet{Be16,Be19}.

In addition to the sample of LyC leakers, we construct a comparison sample of
lower-redshift compact dwarf galaxies with strong emission lines observed
with COS and studied by \citet{Be16,Be19} and observed with STIS and studied
by \citet{Ra20}, totalling 31 galaxies. We extracted {\sl HST} and SDSS
spectra from the respective data bases and measured emission line fluxes and
equivalent widths, similarly to measurements for the LyC leaker sample. Results of the 
measurements are given in Tables \ref{taba1} -- \ref{taba3}.

  \begin{table*}
  \caption{Characteristics for the determination of the C/O abundance ratio in LyC leaking galaxies \label{tab3}}
\begin{tabular}{lcrccrccccccc} \hline
Name        &$F$$^{\rm a}$&EW$^{\rm b}$&$F$$^{\rm a}$&EW$^{\rm b}$&\multicolumn{1}{c}{$F$$^{\rm a}$}&EW$^{\rm b}$&$C_{\rm int}$$^{\rm c}$&$C_{\rm int}$$^{\rm c}$&$C_{\rm int}$$^{\rm c}$&corr.$^{\rm d}$ \\
            &(C~{\sc iii}]&(C~{\sc iii}]&(O~{\sc iii}]&(O~{\sc iii}]&\multicolumn{1}{c}{(H$\beta$}&\multicolumn{1}{c}{(H$\beta$}&(H$\beta$&(C~{\sc iii}]&(O~{\sc iii}]& \\
            &$\lambda$1908)&$\lambda$1908)&$\lambda$1666)&$\lambda$1666)&\multicolumn{1}{c}{$\lambda$4861)}&$\lambda$4861)&$\lambda$4861)&$\lambda$1908)&$\lambda$1666)&\\ \hline
J0901$+$2119&4.5$\pm$1.3&16.7$\pm$5.0 &1.4$\pm$1.2&4.2$\pm$3.6&16.8$\pm$0.9&255$\pm$16&0.22$\pm$0.04& 0.53$\pm$0.10& 0.55$\pm$0.10&1.333\\
J0925$+$1403&7.4$\pm$1.4& 6.4$\pm$1.2 &...        &...        &28.8$\pm$1.1&177$\pm$12&0.21$\pm$0.06& 0.51$\pm$0.15& 0.53$\pm$0.15&1.000\\
J1011$+$1947&6.5$\pm$0.7&18.2$\pm$2.0 &3.7$\pm$1.3&6.7$\pm$2.4&13.9$\pm$0.8&237$\pm$30&0.17$\pm$0.04& 0.41$\pm$0.10& 0.42$\pm$0.10&1.111\\
J1152$+$3400&7.7$\pm$1.2& 4.8$\pm$0.8 &...        &...        &23.3$\pm$0.7&198$\pm$10&0.10$\pm$0.06& 0.24$\pm$0.14& 0.25$\pm$0.15&1.111\\
J1154$+$2443&5.7$\pm$0.7&12.4$\pm$1.5 &4.0$\pm$1.8&7.3$\pm$3.3& 7.7$\pm$0.4&220$\pm$14&0.07$\pm$0.05& 0.17$\pm$0.12& 0.17$\pm$0.12&1.111\\
J1243$+$4646&6.6$\pm$1.0&10.9$\pm$1.7 &1.6$\pm$0.9&2.0$\pm$1.2&11.4$\pm$0.4&221$\pm$10&0.09$\pm$0.04& 0.22$\pm$0.10& 0.22$\pm$0.10&1.000\\
J1248$+$4259&6.8$\pm$0.9&19.0$\pm$2.5 &3.1$\pm$1.1&6.9$\pm$2.4&21.8$\pm$0.6&426$\pm$12&0.22$\pm$0.04& 0.52$\pm$0.09& 0.54$\pm$0.10&1.176\\
J1256$+$4509&4.3$\pm$0.9&14.3$\pm$3.2 &...        &...        & 9.2$\pm$0.4&253$\pm$17&0.10$\pm$0.05& 0.23$\pm$0.11& 0.23$\pm$0.12&1.429\\
J1333$+$6246&4.9$\pm$1.5& 7.0$\pm$2.2 &...        &...        & 9.3$\pm$0.3&194$\pm$13&0.07$\pm$0.09& 0.17$\pm$0.22& 0.17$\pm$0.22&1.282\\
J1442$-$0209&9.5$\pm$1.5& 8.2$\pm$1.3 &4.5$\pm$1.7&3.3$\pm$1.2&29.9$\pm$2.1&312$\pm$12&0.14$\pm$0.06& 0.37$\pm$0.16& 0.38$\pm$0.16&1.111\\
J1503$+$3644&6.4$\pm$1.3& 8.3$\pm$1.6 &...        &...        &19.4$\pm$0.6&297$\pm$12&0.13$\pm$0.07& 0.31$\pm$0.17& 0.32$\pm$0.17&1.250\\
\hline
\end{tabular}

\begin{tabular}{lccccccc} \hline
Name        &$t$(O~{\sc iii})$^{\rm e}$&\multicolumn{4}{c}{$I$($\lambda$)/$I$(H$\beta$)$^{\rm f}$}&log(C/O)$_{\rm opt}$$^{\rm g}$&log(C/O)$_{\rm UV}$$^{\rm h}$\\
            &                &(C~{\sc iii}]&(O~{\sc iii}]&([O~{\sc ii}] &([O~{\sc iii}]&\\ 
            &                &$\lambda$1908)&$\lambda$1666)&$\lambda$3727)&$\lambda$5007)&\\ \hline
J0901$+$2119& 1.22$\pm$0.08  & 0.70$\pm$0.21&0.22$\pm$0.19& 0.82$\pm$0.04& 6.55$\pm$0.26& $-$0.456$\pm$0.291&$-$0.702$\pm$0.410\\
J0925$+$1403& 1.50$\pm$0.09  & 0.49$\pm$0.09& ...         & 1.26$\pm$0.05& 6.08$\pm$0.20& $-$0.925$\pm$0.183& ...    \\
J1011$+$1947& 1.46$\pm$0.08  & 0.69$\pm$0.07&0.40$\pm$0.14& 0.30$\pm$0.02& 8.07$\pm$0.19& $-$0.731$\pm$0.111&$-$0.809$\pm$0.157\\
J1152$+$3400& 1.34$\pm$0.09  & 0.49$\pm$0.08& ...         & 1.06$\pm$0.06& 5.71$\pm$0.26& $-$0.734$\pm$0.215& ...    \\
J1154$+$2443& 1.83$\pm$0.14  & 1.01$\pm$0.12&0.72$\pm$0.32& 0.43$\pm$0.03& 5.78$\pm$0.16& $-$0.768$\pm$0.124&$-$0.900$\pm$0.175\\
J1243$+$4646& 1.57$\pm$0.14  & 0.79$\pm$0.12&0.18$\pm$0.10& 0.54$\pm$0.05& 7.26$\pm$0.20& $-$0.780$\pm$0.256&$-$0.446$\pm$0.361\\
J1248$+$4259& 1.88$\pm$0.12  & 0.70$\pm$0.15&0.36$\pm$0.13& 0.49$\pm$0.03& 5.84$\pm$0.14& $-$0.963$\pm$0.121&$-$0.751$\pm$0.171\\
J1256$+$4509& 1.61$\pm$0.14  & 0.92$\pm$0.19& ...         & 0.44$\pm$0.04& 7.23$\pm$0.21& $-$0.725$\pm$0.215& ...    \\
J1333$+$6246& 1.78$\pm$0.22  & 0.75$\pm$0.23& ...         & 1.30$\pm$0.11& 6.23$\pm$0.41& $-$0.959$\pm$0.281& ...    \\
J1442$-$0209& 1.46$\pm$0.08  & 0.64$\pm$0.10&0.31$\pm$0.12& 0.94$\pm$0.05& 6.24$\pm$0.26& $-$0.760$\pm$0.165&$-$0.841$\pm$0.233\\
J1503$+$3644& 1.49$\pm$0.10  & 0.61$\pm$0.12& ...         & 1.34$\pm$0.08& 6.54$\pm$0.32& $-$0.849$\pm$0.198& ...    \\
\hline
\end{tabular}

\hbox{$^{\rm a}$Observed flux in 10$^{-16}$ erg s$^{-1}$ cm$^{-2}$.}

\hbox{$^{\rm b}$Restframe equivalent width in \AA.}

\hbox{$^{\rm c}$Extinction coefficient for H$\beta$, C~{\sc iii}] $\lambda$1908
and O~{\sc iii}] $\lambda$1666 emission lines.}

\hbox{$^{\rm d}$Correction factor used to adjust observed UV spectrum with the modelled SED.}

  \hbox{$^{\rm e}$$t$(O~{\sc iii}) = $T_{\rm e}$(O~{\sc iii})/10$^4$K, where $T_{\rm e}$(O~{\sc iii}) is the electron temperature in O$^{2+}$ zone derived by the
    direct method.}

\hbox{$^{\rm f}$Extinction-corrected flux ratios.}

\hbox{$^{\rm g}$C/O abundance ratio is derived using C~{\sc iii}] $\lambda$1908 and [O~{\sc iii}] $\lambda$4959,5007 emission-line fluxes.}

\hbox{$^{\rm h}$C/O abundance ratio is derived using C~{\sc iii}] $\lambda$1908 and O~{\sc iii}] $\lambda$1666 emission-line fluxes.}

  \end{table*}

\begin{figure*}
\centering
\includegraphics[angle=0,width=0.99\linewidth]{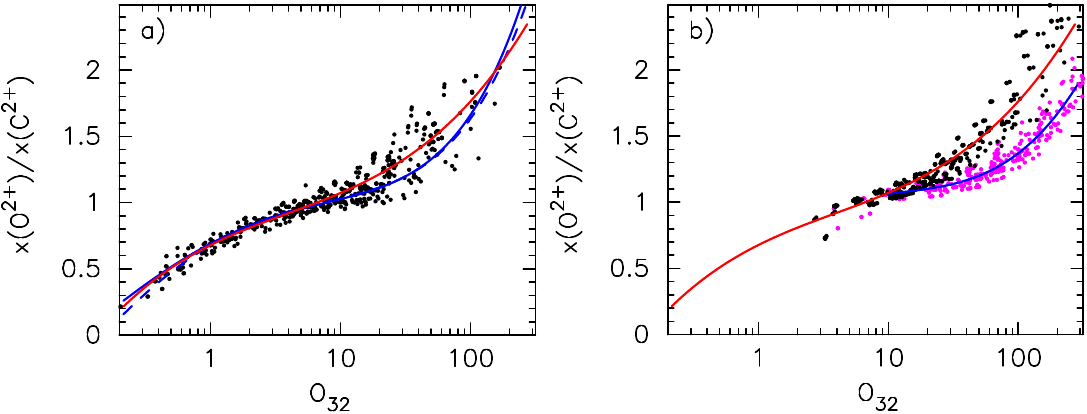}
\caption{{\bf a)} The dependence of the ionization correction factors
$ICF$(C/O) = $x$(O$^{2+}$)/$x$(C$^{2+}$) on the O$_{32}$ =
[O~{\sc iii}]~$\lambda$5007/[O~{\sc ii}]~$\lambda$3727 emission-line ratios
calculated for spherical uniform ionization-bounded H~{\sc ii} regions
with the {\sc cloudy} v. 17.01 code \citep{Fe13}.
Here $x$(O$^{2+}$) and $x$(C$^{2+}$) are volume number fractions of the
respective ions. The red line is the fit to the models (Eq. \ref{eq:icf}).
For comparison, by the blue solid and blue dashed lines, are shown the relations
for $Z$ = 0.1 Z$_\odot$ and 0.2 Z$_\odot$ from \citet{Be19}, respectively.
{\bf b)} The dependence of the ionization correction factors
$ICF$(C/O) on the O$_{32}$ ratio for uniform H~{\sc ii} regions
with a starburst age of 2 Myr, surrounded by an 
envelope of neutral gas with various column number densities $N$(H~{\sc i}). The
models with $N$(H~{\sc i})~$\ge$~10$^{18.5}$ cm$^{-2}$ are represented by black
dots, whereas magenta dots are models with
$N$(H~{\sc i})~$\le$~10$^{18.0}$ cm$^{-2}$, which
are fitted by the blue line. The red line is the same as in {\bf a)}.
\label{fig2}}
\end{figure*}

\begin{figure*}
\includegraphics[angle=0,width=0.99\linewidth]{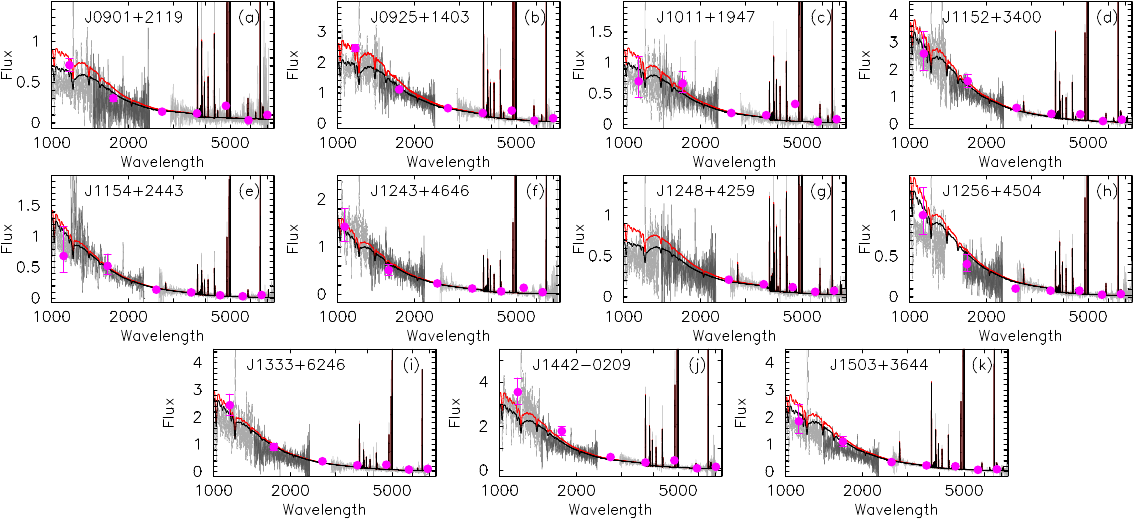}
\caption{The comparison of the observed {\sl HST}/COS (light-grey lines in the
UV range) \citep{I16a,I16b,I18a,I18b}, {\sl HST}/STIS (dark grey lines) and SDSS
(light-grey lines in the optical range) spectra with the SEDs in the
optical range and their extrapolations to the UV range. The extrpolations
are attenuated
adopting the \citet*{C89} reddening law, the derived $C$(H$\beta$) and two
$R(V)$ values of 2.7 (black line) and 3.1 (red line). For comparison,
the photometric fluxes obtained from the {\sl GALEX} FUV, NUV and SDSS
$u,b,r,i,z$ magnitudes are shown by filled magenta circles. All data are
reduced to the rest-frame wavelength scale. 
\label{fig3}}
\includegraphics[angle=0,width=0.99\linewidth]{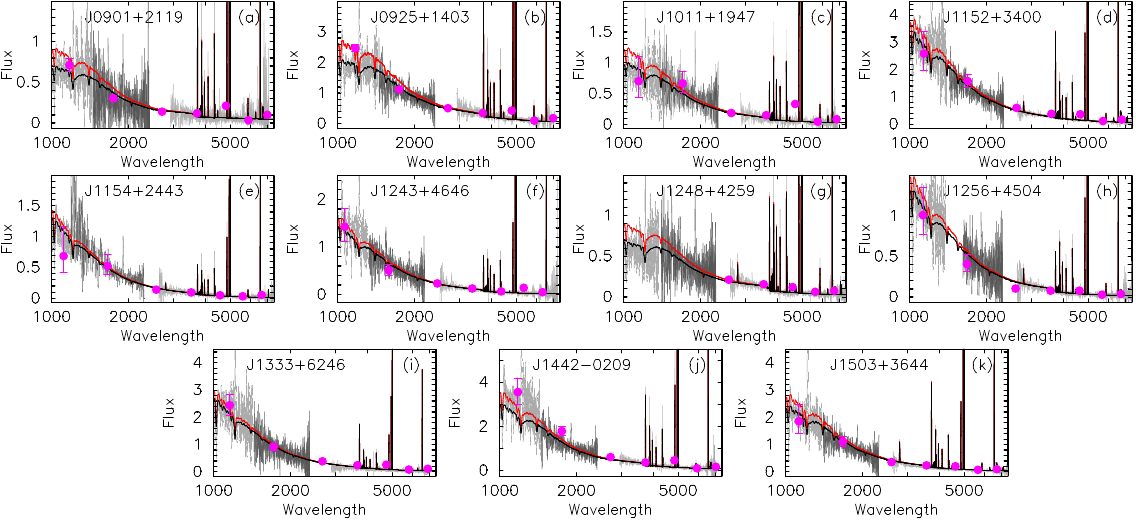}
\caption{The same as in Fig.~\ref{fig3} but with the UV spectra scaled to the
  SEDs with $R$($V$) = 2.7 (black line) at $\lambda$ = 1908\AA. 
\label{fig4}}
\end{figure*}

\begin{figure*}
\includegraphics[angle=0,width=0.99\linewidth]{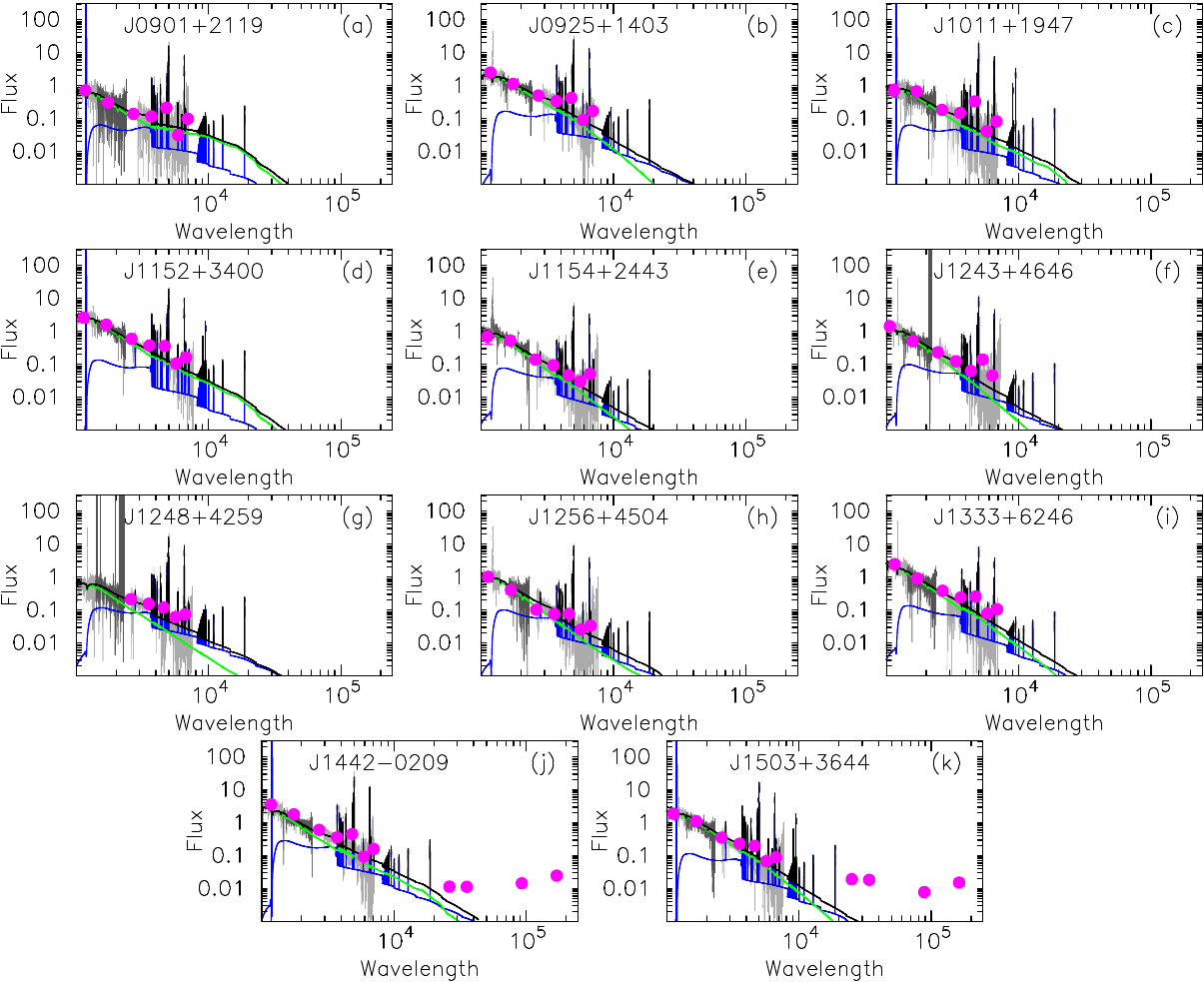}
\caption{The comparison of the observed {\sl HST}/COS (light-grey lines in the
UV range) \citep{I16a,I16b,I18a,I18b}, {\sl HST}/STIS (dark grey lines) and SDSS
(light-grey lines in the optical range) spectra with the SEDs in the
optical range and their extrapolations to the UV and infrared ranges. Stellar
and nebular SEDs are shown by green and blue lines, respectively, and the
total (stellar+nebular) SEDs are represented by black lines.
The photometric fluxes obtained from the {\sl GALEX} FUV, NUV, SDSS
$u,b,r,i,z$ and {\sl WISE} W1 (3.4 $\mu$m), W2 (4.6 $\mu$m), W3 (12 $\mu$m),
W4 (22 $\mu$m) mid-infrared magnitudes are shown by filled magenta circles.
\label{fig5}}
\end{figure*}


\section{The method}\label{sec:method}

We derive CNO element abundances in both the LyC leakage sample and
the comparison sample using the STIS-MAMA \citep[this paper, ][]{Ra20},
COS \citep{Be16,Be19} and SDSS spectra,
corrected for both the Milky Way and internal extinction. The galaxy internal
interstellar extinction has been obtained from the observed decrement
of hydrogen emission lines in the SDSS spectra \citep*{ITL94}
after correcting them for the Milky Way extinction with $A(V)_{\rm MW}$ from the
NED and adopting the \citet*{C89} reddening law with $R(V)_{\rm MW}$~=~3.1.
Second, the fluxes of emission lines at the rest-frame 
wavelengths were corrected for the internal 
extinction of galaxies adopting the \citet{C89} reddening law with
$R(V)_{\rm int}~=~2.7$ and the extinction coefficient $C_{\rm int}$(H$\beta$)
derived from the Balmer decrement, including all observed hydrogen lines
of the Balmer series, after correcting for the Milky Way extinction.

The extinction coefficient is determined by minimizing 
the deviations from the
theoretical case B ratios of all extinction-corrected hydrogen flux ratios. We have checked with the {\sc cloudy} photoionized
H~{\sc ii} models with zero extinction and various column number densities of
the uniform H~{\sc i} envelopes surrounding the H~{\sc ii} region, that the
modelled hydrogen flux ratios are nearly unchanged in models with
$N$(H~{\sc i}) greater than 10$^{17}$ cm$^{-2}$, corresponding to an optical depth
of $\sim$ 1 in the Lyman continuum, and they are consistent with the case B
ratios. Deviations from the case B ratios are seen only at lower
$N$(H~{\sc i}). In particular, the H$\alpha$/H$\beta$ and H$\gamma$/H$\beta$
flux ratios in models with these low column number densities are
respectively lower and higher than the case B ratios. Thus, our correction for
extinction is valid for high H~{\sc i} column densities where the LyC escape fraction is low.

We select  $R(V)_{\rm int}$ = 2.7 because the observed
FUV spectra of LyC leaking galaxies are better fit by extrapolating
the SDSS optical spectra compared to that with a reddening law by \citet{C89}
with a more common $R(V)_{\rm int}$ = 3.1 \citep{I16b,I18b}.

However, the reddening law by \citet{C89} has a NUV dust bump, whereas it is
not seen in galaxies studied in this paper. This bump does not affect the
intensities of the O~{\sc iii}]~$\lambda$1666 emission line, but may slightly
change the intensities of the  C~{\sc iii}]~$\lambda$1908 emission line  in
some galaxies from the LyC leaking sample and some GP galaxies from
\citet{Ra20}. Therefore, we excluded this feature replacing
values of the reddening law in this range with the values interpolated from
the values outside the bump.
Then the extinction-corrected flux $I$ is derived from the observed flux $F$
using the relation $I$ = $F$$\times$10$^{C(\lambda)}$. The extinction coefficients
$C_{\rm int}$(H$\beta$), $C_{\rm int}$(O~{\sc iii}]~$\lambda$1666) and
$C_{\rm int}$(C~{\sc iii}]~$\lambda$1908) for the
H$\beta$, O~{\sc iii}]~$\lambda$1666 and C~{\sc iii}]~$\lambda$1908 emission
lines for the LyC leakage sample are presented in Table~\ref{tab3} and for
the comparison sample in Tables \ref{taba1} -- \ref{taba3}.

Finally, the extinction-corrected emission lines in the optical range
measured in the SDSS spectra are used to derive oxygen and nitrogen ionic and
total abundances following the methods described in \citet{I06}.
As for C$^{2+}$ abundance, we use the relation by \citet{IT99}
\begin{equation}
\frac{{\rm C}^{2+}}{{\rm O}^{2+}}=0.093\exp \left( \frac{4.656}{t}\right)
\frac{I({\rm C~\textsc{iii}]}\lambda 1908)}{I({\rm [O~\textsc{iii}]}\lambda 4959 + \lambda 5007)}
\label{eq:c2po2p}
\end{equation}
to derive the C$^{2+}$/O$^{2+}$ abundance ratio, where $t$ = $T_{\rm e}$/10$^4$K.
We also use the O~{\sc iii}]~$\lambda$1666\AA\ emission line for the
determination of the C$^{2+}$/O$^{2+}$ abundance ratio using
\begin{equation}
\frac{{\rm C}^{2+}}{{\rm O}^{2+}}=0.15\exp \left( \frac{1.1054}{t}\right)
\frac{I({\rm C~\textsc{iii}]}\lambda 1908)}{I({\rm [O~\textsc{iii}]}\lambda 1666)}
\label{eq:c2po2p1}
\end{equation}
\citep{Erb10}. However, the O~{\sc iii}]~$\lambda$1666\AA\ emission line in
spectra shown in Fig.~\ref{fig1} and in galaxies from \citet{Ra20} is weak or
nondetected. Therefore, in the analysis of element abundances we 
mainly use the C/O abundance ratios obtained using strong optical
[O~{\sc iii}] emission lines. On the other hand, the C/O ratios derived from
the C~{\sc iii}] and UV O~{\sc iii}] emission lines are used to check the
consistency of the abundances derived by two methods.

The ratio of total C and O abundances can be derived from the relation
\begin{equation}
\frac{\rm C}{\rm O}=ICF\left(\frac{\rm C}{\rm O}\right)\frac{{\rm C}^{2+}}{{\rm O}^{2+}}, \label{eq:co}
\end{equation}
where the ionization correction factor $ICF$ is determined as a ratio of
O$^{2+}$ and C$^{2+}$ volume number fractions $x$(O$^{2+}$) and $x$(C$^{2+}$) in
the H~{\sc ii} region:
\begin{equation}
ICF\left(\frac{\rm C}{\rm O}\right) = \frac{x({\rm O}^{2+})}{x({\rm C}^{2+})}.
\label{eq:icffit}
\end {equation}

To derive $ICF$(C/O) we use the observable
O$_{32}$\,=\,[O~{\sc iii}]\,$\lambda$5007/[O~{\sc ii}]\,$\lambda$3727, which
is the characteristic of the H~{\sc ii} region ionization state and
depends on the ionization parameter. The O$_{32}$ ratio
can directly be derived from the extinction-corrected SDSS optical spectrum.

We use the {\sc cloudy} v17.01 code \citep{Fe13} to calculate a series of spherical
uniform H~{\sc ii} region models, varying in wide ranges of different parameters such as
metallicity,
production rate of the ionizing radiation $Q$, number density, filling factor,
etc. In our calculations we assume that the escape fraction of ionizing radiation
is zero, i.e. models are ionization-bounded. Therefore, the modelled O$_{32}$ is not affected by the LyC leakage.
Fortunately, LyC escape fractions in most galaxies from our sample are
low, $\la$ 10 per cent. Furthermore, C~{\sc iii}] and [O~{\sc iii}] emission
lines are formed in the inner parts of the H~{\sc ii} region, at variance with
the [O~{\sc ii}] emission, implying that the effect of the leakage on the
ionization correction factor is likely small.
The results of our calculations are shown in Fig.~\ref{fig2}a by black dots.
These data can be fitted by the relation
\begin{equation}
ICF\left(\frac{\rm C}{\rm O}\right) =
0.12295y^3-0.22010y^2+0.49088y+0.67775 \label{eq:icf}
\end{equation}
(red line), where $y$ = log(O$_{32}$). For comparison, the relations between
$ICF$(C/O) and O$_{32}$ obtained by \citet{Be19} for the metallicities
$Z$ = 0.1 Z$_\odot$ and 0.2 Z$_\odot$ are shown by blue lines. All relations are
in very good
agreement for O$_{32}$ $\la$10, but differ somewhat  (by $\la$ 10 per cent
in $ICF$(C/O) at fixed O$_{32}$) for O$_{32}$ $>$ 10. It is seen in
Fig. \ref{fig2}a that the ionization correction factor, and thus the C/O abundance
ratio, can be somewhat overestimated because the O$_{32}$ ratio in these galaxies
may increase due to LyC leakage.

To study the possible impact of LyC leakage on $ICF$, we consider a set of
models with varying column densities $N$(H~{\sc i}) of neutral hydrogen in the
uniform neutral gas envelope surroundng the H~{\sc ii} region. Models with high
and low $N$(H~{\sc i}) correspond to ionization-bounded and density-bounded
H~{\sc ii} regions, respectively. In Fig.~\ref{fig2}b are shown, by black and
magenta dots, models with $N$(H~{\sc i}) $\ge$ 10$^{18.5}$ cm$^{-2}$ and
$\le$ 10$^{18.0}$ cm$^{-2}$,
respectively. It is seen that the models with high $N$(H~{\sc i}) follow the
relation Eq.~{\ref{eq:icf} for ionization-bounded H~{\sc ii} regions (red line),
whereas models with low $N$(H~{\sc i}) are offset to higher O$_{32}$ and their
distribution can be fit by the relation
\begin{equation}
ICF\left(\frac{\rm C}{\rm O}\right) =
0.26863y^3-0.89647y^2+1.11594y+0.57551. \label{eq:icfdb}
\end{equation}
This offset is presumably caused by the reduced intensity of
[O~{\sc ii}]~$\lambda$3727 emission line and thus by the high O$_{32}$ ratio
in the density-bounded H~{\sc ii} regions. Therefore, the use of the relation
Eq.~\ref{eq:icf} for galaxies with high LyC leakage would result in a somewhat
overestimated C/O abundance ratio, by $\la$ 15 per cent, adopting the observed
range of O$_{32}$ $\la$ 15 -- 20. However, the LyC escape fraction is high
($>$ 30 per cent) in only three out of eleven galaxies in our sample.
Furthermore, the LyC escape fraction in the galaxies from the comparison
sample is not known. Therefore, for lack of more information, we adopt the relation
Eq.~\ref{eq:icf} to determine the $ICF$ for all our galaxies.}

The C/O abundance ratios of LyC leaking galaxies and galaxies from the
comparison sample derived by the two methods using UV and optical oxygen
emission lines, respectively, are shown in Table~\ref{tab3}
and Tables \ref{taba1} -- \ref{taba3}, respectively.
The ratios derived by the method using optical [O~{\sc iii}] emission
lines can be subject to differences in spectroscopic apertures used in the UV
and optical ranges. However, many of the selected galaxies are very compact,
therefore the effect of different
apertures is likely small. The spectra can also be subject to uncertainties in
data reduction and absolute flux calibration. Specifically, for spectrographs
using MAMA detectors in the UV (both the STIS and COS) some data can be lost
if the buffer time for saving the data is non optimally selected. We assume that
the derived C/O abundance ratios are reliable if the observed STIS (or COS)
spectrum is consistent with the extrapolation of the attenuated SED, which is
modelled from the SDSS spectrum. This requirement implies that emission in both
the UV and optical ranges is dominated only by radiation of the same massive
stars. This is likely the case, because equivalent widths EW(H$\beta$)
in spectra of all galaxies considered here are high, $>$ 100\AA\
(Table \ref{tab3}, Tables
\ref{taba1} -- \ref{taba3}).

\begin{figure}
\includegraphics[angle=-90,width=0.99\linewidth]{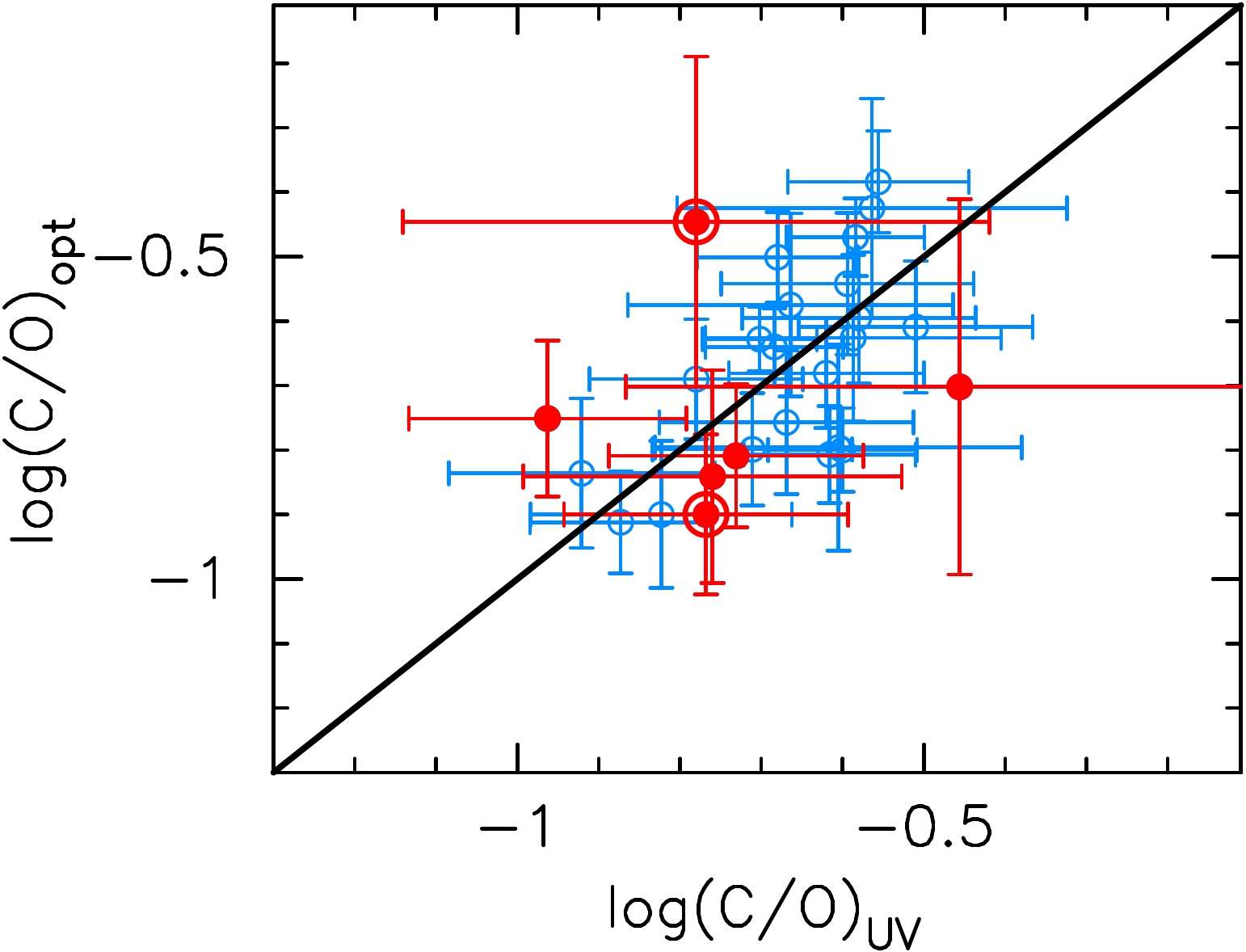}
\caption{Comparison of the carbon-to-oxygen abundance ratios derived using
O~{\sc iii}]\,$\lambda$1666 ((C/O)$_{\rm UV}$) and
[O~{\sc iii}]\,$\lambda$4959, 5007
((C/O)$_{\rm opt}$) emission lines. Galaxies from \citet{Be16,Be19} and
this paper are shown by blue open circles and filled red circles, respectively.
Encircled symbols indicate the galaxies with $f_{\rm esc}$(LyC)\,$>$\,30 per cent.
The solid line is the line of equal values.}
\label{fig6}
\end{figure}

\begin{figure*}
\includegraphics[angle=0,width=0.99\linewidth]{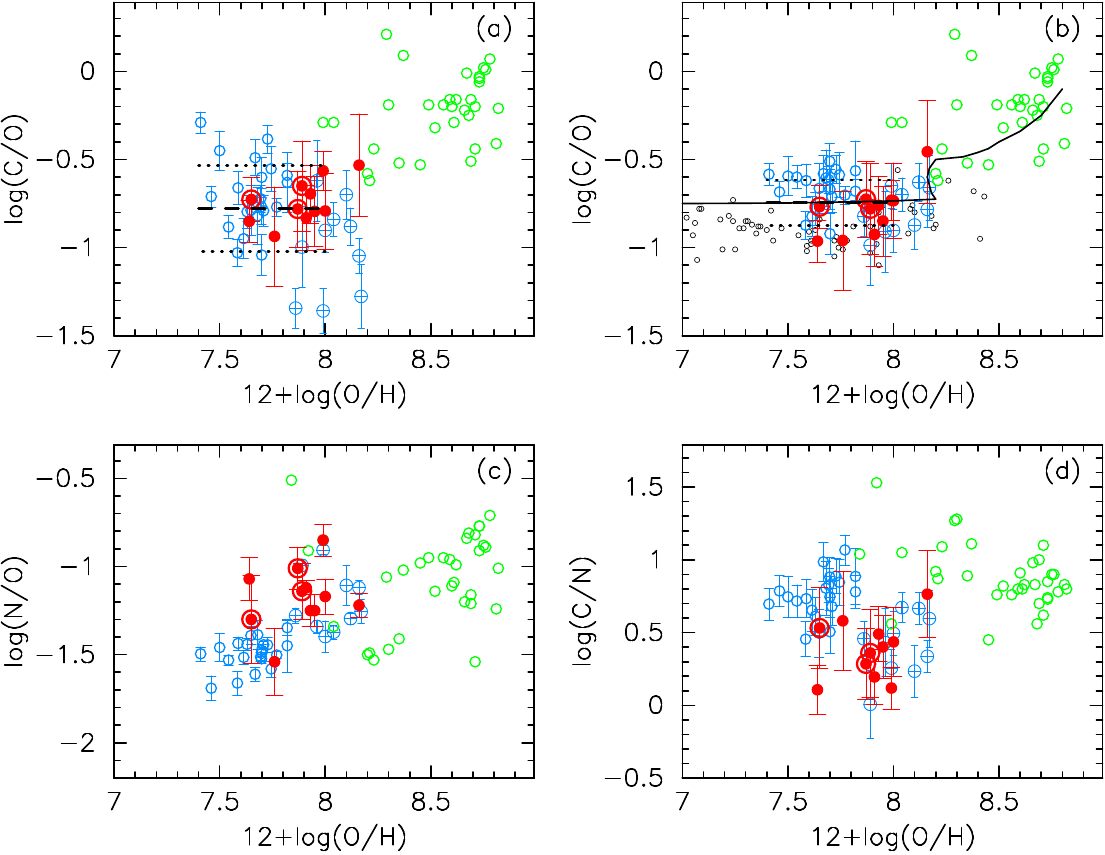}
\caption{{\bf (a)} The relation between the oxygen abundance 12 + log(O/H)
and log(C/O). Carbon abundances are derived from the
extinction-corrected C~{\sc iii}]\,$\lambda$1908\AA\ emission line (red and
blue symbols) and from the extinction-corrected recombination line
C~{\sc ii}\,$\lambda$4267\AA\ (green symbols). No corrections to
adjust UV and optical spectra were applied for galaxies, shown by blue
symbols. {\bf (b)} The same as in {\bf (a)}, but there is an adjustment to the
UV spectra such that the optical and UV fluxes agree with SED models.
{\bf (c)} The relation 12 + log(O/H) vs log(N/O). {\bf (d)} The relation
12 + log(O/H) vs log(C/N). The correction to adjust UV and optical spectra
is applied. In all panels filled red circles are from this paper,
open blue circles are from \citet{Be16,Be19},
crossed open circles are from \citet{Ra20},
and green symbols are from \citet{Es02,Es09,Es14}, \citet{Ga07}, \citet{Lo07}.
Small black open circles in {\bf (b)} are halo stars \citep{Ak04,Fa09}, black
line is the chemical evolution model by \citet{Ca11}. Encircled symbols
indicate the galaxies with $f_{\rm esc}$(LyC) $>$ 30 per cent. Dashed lines in
{\bf (a)} and {\bf (b)} indicate average log(C/O) for the galaxies with the
oxygen abundances in the range 7.4 -- 8.0, whereas dotted lines are 1$\sigma$
deviations from the average values.
\label{fig7}}
\end{figure*}

\begin{figure*}
\includegraphics[angle=0,width=0.99\linewidth]{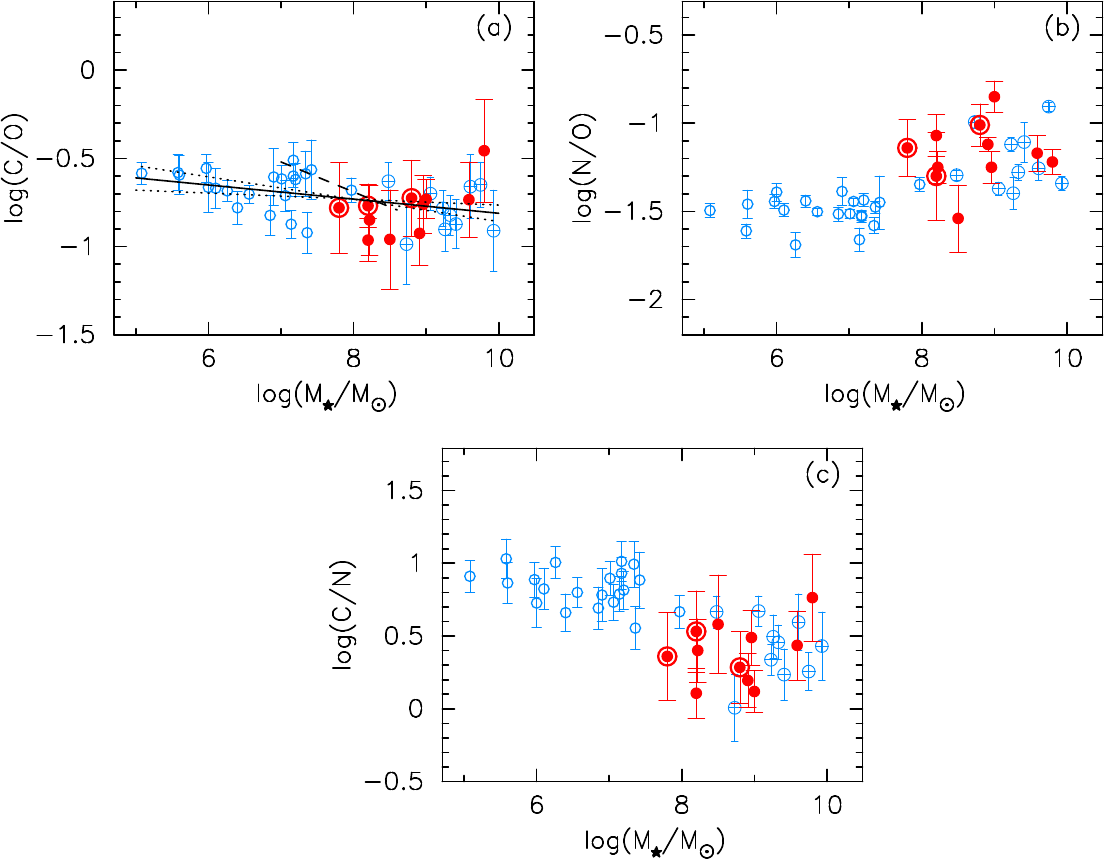}
\caption{Relations between log($M_\star$/M$_\odot$) and {\bf (a)} scaled log(C/O),
{\bf (b)} log(N/O) and {\bf (c)} log(C/N). In {\bf (a)} the most likelihood
relation (Eq.~\ref{eq:coamt}) with 1$\sigma$ alternatives are shown by solid
and dotted lines, respectively, whereas the relation between log(C/O) and
the baryonic mass by \citet{Be19} is represented with a dashed line. Symbols
are the same as in Fig.~\ref{fig7}.
\label{fig8}}
\end{figure*}

We use the SDSS spectra of our LyC leakers and galaxies from a comparison sample
to fit the SEDs. These SEDs are also
used to derive the stellar masses. Briefly, the fitting method,
using a two-component model (young instantaneous burst with an age
$<$~10~Myr and an older continuous star formation at ages $>$~10~Myr), is described
in \citet{I18a,I18b}. The model includes a nebular continuum emission. Its
contribution is determined by the ratio of the observed H$\beta$ equivalent
width EW(H$\beta$) to the value of $\sim$~900~--~1100\AA\ expected for the pure
nebular recombination emission for the range of the electron temperature
$T_{\rm e}$~=~10000~--~20000K \citep[e.g. ][]{A84}.
A $\chi ^2$ minimisation technique was used to fit the continuum 
in the rest-frame wavelength range $\sim$~3000~--~6500\AA\ and to reproduce 
the observed H$\beta$ and H$\alpha$ equivalent widths. Inclusion of
EW(H$\beta$) and EW(H$\alpha$) in minimisation is critically important because
they constrain the age of the starburst and flux of the ionizing radiation
shortward 912\AA\ and thus the modelled continuum in the UV range at longer
wavelengths.

In Fig. \ref{fig3} we present the modelled stellar SEDs for the LyC leakers
attenuated with an extinction coefficients $C$(H$\beta$) derived from the
Balmer decrement (Table~\ref{tab3}) and \citet{C89}
reddening laws with two values of $R(V)$ = 2.7 (black lines)
and 3.1 (red lines).
They, together with the fluxes derived
from the {\sl GALEX} FUV and NUV magnitudes
and SDSS $u,g,r,i,z$ magnitudes (magenta filled circles), are superposed upon
both the rest-frame COS
\citep[light-grey line in the UV range, ][]{I16a,I16b,I18a,I18b}, STIS
(dark-grey line, this paper) and SDSS (light-grey line in the optical range)
spectra. For all galaxies we find that the modelled SEDs follow the continua
of the COS/FUV and STIS/NUV spectra fairly well, implying
that there is no considerable effect of different apertures or offsets which
may be caused by the inproper buffer time and absolute flux calibration.
The photometric and spectroscopic data are also in fair agreement. This gives us
confidence that the UV fluxes of our LyC leaking galaxies are
robust and are not affected much by the effects discussed above. But still some
small offsets between the observed UV spectra and SEDs are present for some
galaxies. These offsets are considerably higher in the FUV range, than in the
NUV range. Therefore, we introduce scaling factors (``corr'' in
Table~\ref{tab3}) to adjust the observed UV spectra to the attenuated SEDs
at the rest-frame wavelength 1908\AA. For this we adopted the attenuated
SEDs with $R(V)$ = 2.7 (black lines) which better fit the COS FUV and NUV
observed spectra compared to other attenuated SEDs in Fig.~\ref{fig3}, and
multiply observed UV spectra by these scaling factors. The adjusted spectra are
represented in Fig.~\ref{fig4} and they show good coincidence with the
attenuated SEDs (black lines).

The spectra of galaxies from the comparison sample were similarly adjusted with
the SEDs (Figs.~\ref{figa1} - \ref{figa6}) using scaling factors given in
Tables \ref{taba1} -- \ref{taba3}.

\begin{figure}
\centering
\includegraphics[angle=-90,width=0.99\linewidth]{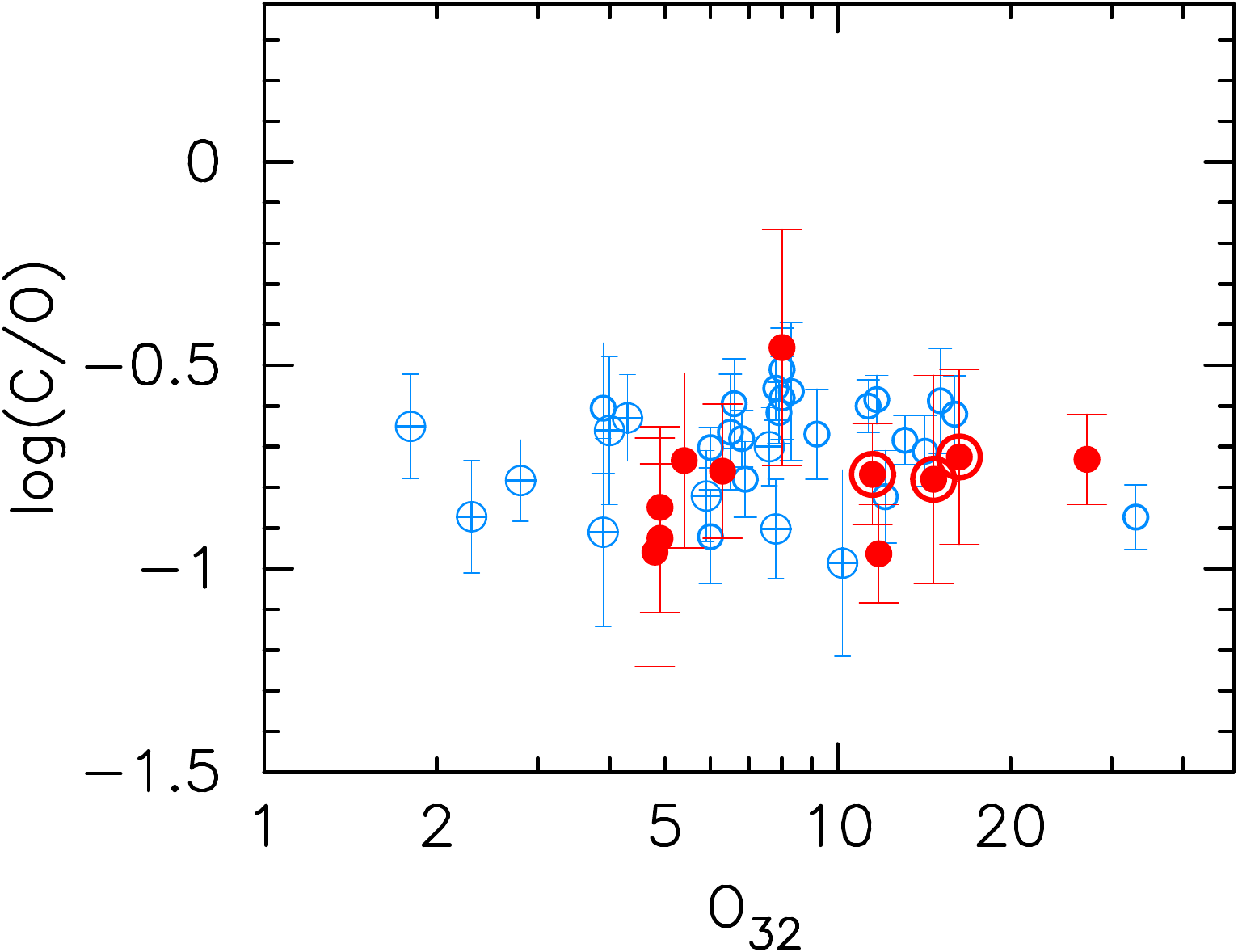}
\caption{The relation between log(C/O), derived from C~{\sc iii}]\,$\lambda$1908
and [O~{\sc iii}]\,$\lambda$4959, 5007 emission-line fluxes, and
O$_{32}$\,=\,[O~{\sc iii}]\,$\lambda$5007/[O~{\sc ii}]\,$\lambda$3727.
Symbols are the same as in Fig.~\ref{fig7}.
\label{fig9}}
\end{figure}

We note, however, that the stellar mass derived from our SEDs may be
somewhat underestimated because the infrared range is not included in the
fitting. Unfortunately, both the galaxies from our and the comparison sample are
faint, and none of them is included in the 2MASS catalogue. However, some of the
brighter galaxies are present in the AllWISE catalogue, at the longer wavelengths
of 3.4, 4.6, 12, 22 $\mu$m. It has been shown, e.g. in \citet{Iz14}, that for most
compact star-forming galaxies the 3.4 $\mu$m -- 4.6 $\mu$m colour in the range
0 -- 0.5 mag corresponds to stellar and nebular emission, whereas the
contribution of dust emission becomes important at longer wavelengths. However,
in a small fraction of compact star-forming galaxies, the
3.4 $\mu$m -- 4.6 $\mu$m colour can be much higher, attaining values greater than
2 mag in
some rare cases \citep{Gr11,Iz11,Iz14}. This implies that the contribution of
hot dust emission at 3.4 and 4.6 $\mu$m in these galaxies is important and
even dominant.

In Fig.~\ref{fig5}, we present the spectral energy distribution of our
galaxies including near- and mid-infrared ranges. Only two galaxies were
detected by the {\sl Wide-field Infrared Survey Explorer} ({\sl WISE}).
On the other hand, many galaxies from the comparison
sample are detected by {\sl WISE}, mainly because of their higher brightness
(Fig.~\ref{figa7} -- \ref{figa9}). The 3.4 $\mu$m and 4.6 $\mu$m fluxes
in many galaxies from the comparison sample are consistent with the predictions
of our modelled SEDs, if aperture corrections are taken into account. This
implies that the contribution of emission from cool stars is small, indicating
a relatively small mass of their population, at least, inside the spectroscopic
aperture. An indication of hot dust emission is present in some galaxies from
the comparison sample, most notably in J120122 (Fig.~\ref{figa7}c),
J2238$+$1400 (Fig.~\ref{figa8}m), J030321$-$075923 (Fig.~\ref{figa9}a).
We also note that the contribution of nebular continuum in the infrared
range is high in all galaxies considered in this paper, most notably in
J1248+4259 (Fig.~\ref{fig5}g), where the nebular continuum starts dominating 
from $\sim$ 5000\AA. Additionally, the presence of strong hydrogen and helium
emission lines with very high equivalent widths in the near-infrared range
will further alter the $J,H,K$ magnitudes from those of the pure stellar
emission, making difficult a stellar mass determination with the use
of the near-infrared data.

In Fig.~\ref{fig6} we compare the carbon-to-oxygen abundance ratios derived
from the matched spectra using the O~{\sc iii}]~$\lambda$1666 ((C/O)$_{\rm UV}$))
and [O~{\sc iii}]~$\lambda$4959, 5007 ((C/O)$_{\rm opt}$)) emission lines. The
line O~{\sc iii}]~$\lambda$1666 is detected in all galaxies from
\citet{Be16,Be19}, in 6 LyC leaking galaxies, but almost not seen in
the GP spectra from \citet{Ra20} and therefore they were not used for a
comparison. The figure shows a good agreement between the C/O abundance
ratios derived by the two methods. This supports the reliability of the
C/O abundance ratios derived using [O~{\sc iii}]~$\lambda$4959, 5007
emission lines.

\section{Relations between abundances of CNO elements}\label{sec:abund}

To study the relations between the C/O, N/O and C/N abundance ratios
with oxygen abundances 12~+~logO/H and stellar masses $M_\star$ we use a
sample of 42 galaxies including 11 LyC leaking galaxies at redshifts
$z$~$\sim$~0.3~--~0.4 (this paper), 10 GP galaxies at redshifts
$z$~$\sim$~0.15~--~0.25 \citep{Ra20} and 21 galaxies at lower redshifts
$z$~$\la$~0.04 \citep{Be16,Be19}. All measurements of emission-line fluxes
and determinations of element abundances have been done
in this paper. In particular, we used the C/O abundance ratios derived from the
C~{\sc iii}]~$\lambda$1908 and [O~{\sc iii}]~$\lambda$4959, 5007 emission-line
fluxes (Eq.~\ref{eq:c2po2p}).

In Fig.~\ref{fig7}a we show the distribution in the
12 + log(O/H)~--~log(C/O)
diagram of the LyC leaking galaxies (red symbols) and of the comparison sample
(blue symbols). Here C/O abundance ratios are derived using non-scaled UV
spectra. The LyC leaking galaxies J1154$+$2443, J1243$+$4646 and
J1256$+$4509 with the escape fraction $f_{\rm esc}$(LyC)~$>$~30 per cent
(Table~\ref{tab1}) are encircled. The green symbols in the figure also show the
data for H~{\sc ii} regions in our Galaxy \citep{Es02,Es09,Es14,Ga07,Lo07}, for
which abundances were derived from the optical recombination carbon and oxygen
lines. It is seen that there is no evident dependence of log(C/O) on 12\,+\,log(O/H)
for the galaxies shown by red and blue symbols which have an average value
$-$0.83\,$\pm$\,0.25, calculated in the range 12\,+\,log(O/H)\,=\,7.4\,--\,8.0 (dashed line)
with 1$\sigma$ dispersions (dotted lines). The derived average value is close to
that obtained by \citet{Be16,Be19}.

We suggest that the
large dispersion for galaxies in Fig.~\ref{fig7}a is due to that
the UV and optical spectra of these galaxies are not well adjusted,
i.e. the energy distribution in the UV spectra can not be reproduced by the
SEDs derived from the SDSS spectra (Figs.~\ref{fig3}, \ref{figa1}, \ref{figa3},
\ref{figa5}).

The comparisons of the observed UV and optical spectra and photometric data with
the modelled SEDs are presented in Figs.~\ref{figa1},  \ref{figa3}, \ref{figa5}.
Comparing these figures with Fig.~\ref{fig3} we note that photometric fluxes
(magenta filled circles) from galaxies studied by \citet{Be16,Be19}
are systematicaly higher than the spectroscopic fluxes in both the UV and
optical ranges, indicating that angular sizes of these galaxies are
larger than spectroscopic apertures. The agreement is better for the GP galaxies
at higher redshifts from the \citet{Ra20} sample and is good for LyC leaking
galaxies (Fig.~\ref{fig3}).

Another feature is that the modelled SEDs fail to reproduce the energy
distribution in the UV range in many galaxies from the \citet{Be16,Be19}
and \citet{Ra20} samples. The agreement is better for LyC leaking
galaxies (Fig.~\ref{fig3}). To overcome this problem we scaled
UV spectra to adjust them with the attenuated SEDs. The UV emission line intensities are also changed by the same factor.

Using scaled intensities in the optical range, we calculate log(C/O) in
LyC leakers and in galaxies from the \citet{Be16,Be19} and \citet{Ra20} samples.
Results of this correction are shown in Fig.~\ref{fig7}b on the
log(C/O)\,--\,12\,+\,log(O/H) diagram. It is seen that the distribution of the
galaxies is much tighter compared to that  in Fig.~\ref{fig7}a and
all data are in much better agreement, showing no dependence of log(C/O) on
the oxygen abundance at these low 12\,+\,log(O/H) values. This indicates a common
carbon and oxygen origin. Although the average value of log(C/O)\,=\,$-$0.75
for galaxies with the oxygen abundances in the range 7.4\,--\,8.0
from the \citet{Be16,Be19}
and \citet{Ra20} samples
(dashed line) is similar to that in Fig.~\ref{fig7}a, the dispersion is
considerably reduced to the value of 0.13 dex. The distribution of the galaxies
with modified log(C/O) values is in good agreement with the chemical evolution
model by \citet{Ca11} and, for \citet{Be16,Be19} data, above by
0.20 dex, compared to that for the galactic halo stars from \citet{Ak04} and \citet{Fa09}.
On the other hand, the distribution of LyC leaking galaxies (red symbols) and
GPs by \citet{Ra20} (blue crossed open circles) are in better agreement with the
distribution of the halo stars.

The distribution of log(N/O) on oxygen abundance 12\,+\,log(O/H)
is shown in Fig.~\ref{fig7}c. Higher-redshift LyC leaking galaxies (this paper)
and GPs by \citet{Ra20} have systematically higher, by a factor of $\sim$ 2,
N/O abundance ratios compared to lower-redshift galaxies from the
\citet{Be16,Be19} samples.
The latter galaxies likely do not show a
dependence of the N/O ratio on oxygen abundance indicating a common origin of
N and O at these metallicities.
On the other hand, in LyC leakers (this paper) and GPs \citep{Ra20}, likely,
an additional source of nitrogen nucleosynthesis is required, possibly,
stellar winds from the most massive stars.

Finally, the distribution of galaxies in the log(C/N) -- 12\,+\,log(O/H) diagram
is shown in Fig.~\ref{fig7}d. LyC leakers and GPs have considerably lower
C/N abundance ratios compared to nearly constant values for galaxies at lower
redshift and H~{\sc ii} regions in the Milky Way. This appearance is presumably
due to higher nitrogen abundances in LyC leakers and GPs, again favouring
an additional mechanism of nitrogen production in higher-redshift galaxies.

We present in Fig.~\ref{fig8} the dependencies on the galaxy stellar mass
$M_\star$ of scaled C/O, N/O and C/N abundance ratios. The log(C/O)
slightly declines with increasing $M_\star$ (Fig.~\ref{fig8}a) and this decline
can be approximated by the maximum likelihood relation
\begin{equation}
\log \left(\frac{\rm C}{\rm O}\right) = -(0.040\pm 0.023)\times \log \left(\frac{M_\star}{{\rm M}_\odot}\right)-(0.410\pm 0.185). \label{eq:coamt}
\end{equation}
This relation together with 1$\sigma$ alternative relations are shown in
Fig.~\ref{fig8}a
by solid and dotted lines, respectively. Previously a similar trend with a higher
slope for a smaller sample in a smaller range of masses 
was noted by \citet{Be19} from the log C/O -- total baryonic mass diagram.
It is shown in Fig.~\ref{fig8}a by a dashed line.
However, we note that a total baryonic mass relationship is not completely
comparable to the stellar mass relationship.

We check whether the trend in Fig.~\ref{fig8}a can be caused by the systematics
in the determination of the carbon abundance. In particular, are the ionization
correction factors (Fig.~\ref{fig2}, Eq.~\ref{eq:icf}) reliably take into
account all stages of carbon ionization? For this, we consider a diagram
log(C/O) -- O$_{32}$ (Fig.~\ref{fig9}). There is no trend in this diagram,
which implies that the applied ICF(C/O) values are correct.
We also note that the C/O
abundance ratio in galaxies
with $f_{\rm esc}$(LyC) $>$ 30 per cent is not systematically different from that
in other LyC leaking galaxies with lower $f_{\rm esc}$(LyC) and in galaxies from
the comparison sample (see also Fig~\ref{fig7}b).

The trend in Fig.~\ref{fig8}a can also be attributed to a decreasing C and O
effective yield with increasing stellar mass. More massive
galaxies with $M_\star$ $\ga$ 10$^8$ M$_\odot$, including LyC leaking galaxies
(red symbols) and GPs by \citet{Ra20} (blue crossed open circles)
have lower log(C/O) by $\sim$ 0.2 dex compared to the average value
for lower-mass galaxies by \citet{Be16,Be19}.

The log(N/O) for low-mass galaxies by \citet{Be16,Be19}
is nearly constant, with an average value of $\sim$ $-$1.5 (Fig.~\ref{fig8}b).
On the other hand, the N/O ratio in higher-mass LyC leakers and GPs is not flat
and enhanced by a factor of $\sim$ 2, with an average log(N/O) $\sim$ $-$1.2.
Similar enhanced log(N/O) was found in $z$ = 2 -- 3 star-forming galaxies
with similar stellar masses, oxygen abundances and excitation characteristics
of their H~{\sc ii} regions (strength of emission lines and O$_{32}$)
\citep{St14,Sh15,St17}.
Correspondingly, the log(C/N) in higher-mass LyC leakers and GPs is
$\sim$ 0.5 dex lower than in low-mass galaxies from \citet{Be16,Be19}.
The transition in properties from low- to high-mass galaxies
occurs at $M_\star$ $\sim$ 10$^8$M$_\odot$ (Fig.~\ref{fig8}c).

In the light of findings in this paper, a successful model of carbon, nitrogen
and oxygen origin in dwarf star-forming galaxies should reproduce all trends of
C/O, N/O and C/N abundance ratios with oxygen abundance and stellar mass,
namely, 1) the constancy of the C/O abundance ratio (Fig.~\ref{fig7}b) with oxygen
abundance, the N/O jump up and the C/N jump down at 12 + log(O/H) $\ga$ 7.8
(Figs.~\ref{fig7}c and \ref{fig7}d), and
2) the steady decrease of the C/O abundance ratio with $M_\star$ (Fig.~\ref{fig8}a),
the N/O jump up and C/N jump down at $M_\star$ $\ga$ 10$^8$ M$_\odot$
(Figs.~\ref{fig8}b and \ref{fig8}c).

Various mechanisms of the C/O and N/O evolution were considered in the past.
In general, oxygen is produced by massive stars whereas carbon is a primary
element produced by massive and intermediate-mass stars. Nitrogen is
produced mainly in the intermediate-mass stars, as a primary element at low
metallicities and as a secondary element at high metallicities. For example,
\citet{H00} have shown that both C/O and N/O abundance ratios
steadily increase with oxygen
abundance. It was shown by e.g. \citet{Pi92,Pi93}, \citet{H00}, \citet*{Yi11},
\citet{Be16,Be19} that bursty models in dwarf galaxies (either
closed-box or with the loss of synthesized oxygen via galactic winds) are able 
to reproduce the temporal evolution and the spread of
C/O and N/O abundance ratios. The nitrogen enhancement via local pollution of
the ISM by massive O and WR stars was also considered
\citep[e.g. ][]{H00,I06}. In particular, \citet{I06} have suggested that the
observed high N/O abundance ratio is not characteristic of the entire
H~{\sc ii} region, but rather of dense nitrogen-enriched clumps expelled by
Wolf-Rayet stars. In general, in all these models both the C/O and N/O
abundance ratios increase with increasing 12\,+\,log(O/H), which does not exactly agree with our data.

The relation $M_\star$ -- log(C/O) in Fig.~\ref{fig8}a, at least in part, may
be caused by decreasing efficiency of the galactic wind expelling oxygen
from the galaxy in higher-mass galaxies with higher gravitational potentials.

Our data indicate that there is an additional source of nitrogen enrichment,
possibly stellar winds from the most massive stars in more massive galaxies.
Some indication of this is found by \citet{I18b} who noted that strong
O~{\sc vi} $\lambda$1035 stellar line
with P Cygni profile is detected in almost all LyC leakers shown by red
symbols. This implies that massive stars with masses above
100~M$_\odot$ are present in these galaxies. On the other hand, the O~{\sc vi}
line is almost not seen in galaxies with $M_\star$ $<$ 10$^8$ M$_\odot$
\citep{I21} with similarly young starburst ages.
Thus, likely, the
IMF in more massive galaxies extends to stars with higher masses, which
more likely can go through the phase with stellar wind, including the
Wolf-Rayet stage, and enrich the interstellar medium with nitrogen.
Unfortunately, we can not prove this statement, because the redshifts of
low-mass galaxies from the sample by \citet{Be16,Be19} are too low for
observation of the O~{\sc vi} $\lambda$1035 at present.

\section{Conclusions}\label{sec:summary}

We present new {\sl Hubble Space Telescope} ({\sl HST})/Space Telescope Imaging
Spectrograph (STIS) observations of eleven $z$ $\sim$ 0.3 -- 0.4 Lyman
continuum (LyC) leaking galaxies from the sample by \citet{I16a,I16b,I18a,I18b}.
We use these data in combination with optical Sloan Digital Sky Survey (SDSS)
observations to study the abundances of CNO elements and relations between them.
These data were supplemented by the archival {\sl HST} STIS and Cosmic Origins
Spectrograph (COS) observations of the C~{\sc iii}] $\lambda$1908\AA\ emission line,
together with available SDSS spectra of low-redshift dwarf galaxies, totalling
in a sample of 42 galaxies. Our main results are summarised as follows:

1. The continua of all 11 LyC-leaking galaxies in the UV spectra can be 
reproduced by extrapolation of the spectral energy distributions (SEDs)
which are modelled using the SDSS optical spectra. This implies that a) the spectra
in the UV and optical ranges are well adjusted. Their fluxes are in agreement
with {\sl Galaxy Evolution Explorer} ({\sl GALEX}) and SDSS photometric fluxes and therefore only small aperture
corrections are needed to compare data in these ranges and b) emission in the UV
and optical ranges is almost totally produced by the stars formed in the same
young burst of star formation.

2. For many other galaxies selected from the literature, the UV and optical
spectroscopic fluxes are considerably lower than the fluxes obtained from the
{\sl GALEX} and SDSS photometry, meaning that the angular sizes of galaxies
are larger than the spectroscopic apertures in both the UV and optical ranges.
Furthermore, in many of these galaxies the continua in the UV range cannot
be reproduced by extrapolation of the SED modelled for SDSS spectra. In these
cases we adopt scaling factors in order to fit both the UV and optical spectra
by the same SED modelled for the SDSS spectrum, assuming that emission in both
ranges is produced by the stars formed during the same young burst of star
formation.

3. Both the UV and optical data are corrected for extinction which is derived
from the hydrogen Balmer decrement. The carbon abundance is derived from the
C~{\sc iii}] $\lambda$1908\AA\ emission line measured in the STIS or COS
spectra, whereas nitrogen and oxygen abundances are obtained from the
[N~{\sc ii}]\,$\lambda$6584\AA\ and [O~{\sc ii}]\,$\lambda$3727\AA,
[O~{\sc iii}]\,$\lambda$4959\AA, $\lambda$5007\AA\ emission lines, respectively,
measured in the SDSS spectra. Adopting the method of SED fitting that
consistently reproduces the UV and optical spectra, we considerably reduce (by
a factor of $\sim$\,2) the dispersion in log(C/O). We find that log(C/O) at
low oxygen abundances 12\,+\,log(O/H) $\la$\,8.0 is constant with an average value
of $-$0.75\,$\pm$\,0.13 for the total sample of selected galaxies.
This implies a common origin of oxygen and carbon. The LyC leakers and
``green peas'' (GP) studied by \citet{Ra20} are shifted 
in the log(C/O) -- 12\,+\,log(O/H) diagram to lower values of log(C/O)
by $\sim$\,0.10\,--\,0.15 dex. The N/O abundance ratio in LyC leakers and GPs
is $\sim$\,2 higher than that of the rest of the sample. This results in
a factor of $\sim$\,3 lower C/N ratio in the LyC leakers and GPs compared to
other selected galaxies.

4. We find a trend of decreasing log(C/O) with increasing stellar mass $M_\star$.
In this relation, the LyC leakers and GPs are located at the high-mass end of
the diagram with $M_\star$ $\ga$ 10$^8$ M$_\odot$.
A similar trend, but with a higher slope, has been obtained by \citet{Be19}
for a smaller sample with a smaller range of masses. The N/O abundance
ratio in low-mass galaxies with $M_\star$ $\la$ 10$^8$ M$_\odot$ is constant
with log(N/O) $\sim$ $-$1.5, indicating a primary origin for nitrogen.
The N/O abundance ratio in higher-mass LyC leakers and GPs is enhanced by a
factor of $\sim$ 2 to higher values, indicating an additional source of
nitrogen enrichment.

5. We speculate that the observed abundance ratios and the transition of 
N/O abundance ratio from low values in low-mass galaxies to higher 
values at $\ga$ 10$^8$ M$_\odot$ could be due to different stellar IMFs which 
extend to higher masses of massive stars in higher-mass galaxies.
We also discuss alternate explanations and conclude that the origin of
the enhanced N/O abundances in some of LyC leakers \citep[cf. ][]{Gu20} and GPs
remains unexplained.

\section*{Acknowledgements}

Based on observations made with the NASA/ESA {\sl Hubble Space Telescope}, 
obtained from the data archive at the Space Telescope Science Institute. 
STScI is operated by the Association of Universities for Research in Astronomy,
Inc. under National Aeronautics and Space Administration (NASA) contract NAS 5-26555. Support for this work was provided by 
NASA through grant number HST-GO-15941.002-A from the Space Telescope Science 
Institute, which is operated by Association of Universities for Research in
Astronomy (AURA), Inc., under NASA contract NAS 5-26555.
Y.I. acknowledges support from the National Academy of Sciences of 
Ukraine (Project ``The matter properties at high energies and in galaxies
during the reionization of the Universe'') and from the Simons Foundation.
 Funding for Sloan Digital Sky Survey-III (SDSS-III) has been provided by the Alfred P. Sloan Foundation, 
the Participating Institutions, the National Science Foundation, and the U.S. 
Department of Energy Office of Science. The SDSS-III web site is 
http://www.sdss3.org/. SDSS-III is managed by the Astrophysical Research 
Consortium for the Participating Institutions of the SDSS-III Collaboration. 
{\sl Galaxy Evolution Explorer} ({\sl GALEX}) is a NASA mission  managed  by  the  Jet  Propulsion  Laboratory.
This research has made use of the NASA/Infrared Processing and Analysis Center
(IPAC) Extragalactic Database (NED) which 
is operated by the Jet  Propulsion  Laboratory,  California  Institute  of  
Technology,  under  contract with the National Aeronautics and Space 
Administration. This publication makes use of data products from the
{\sl Wide-field Infrared Survey Explorer} ({\sl WISE}), which is a joint project of the
University of California, Los Angeles, and the Jet Propulsion Laboratory,
California Institute of Technology, funded by the National Aeronautics and
Space Administration.

\section*{Data availability}

The data underlying this article will be shared on reasonable request to the 
corresponding author.







\appendix

\section{SED fits of galaxies spectra from the literature}




\begin{figure*}
\includegraphics[angle=0,width=0.99\linewidth]{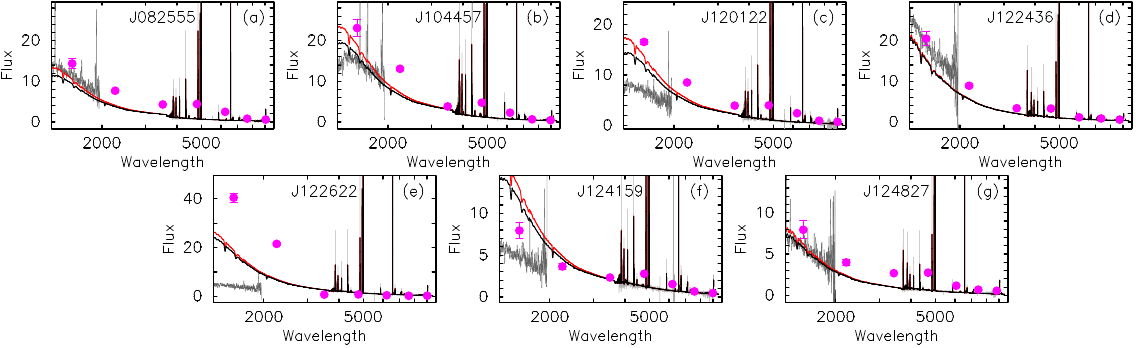}
\caption{The same as in Fig.~\ref{fig3} but for galaxies from \citet{Be16}. 
\label{figa1}}
\includegraphics[angle=0,width=0.99\linewidth]{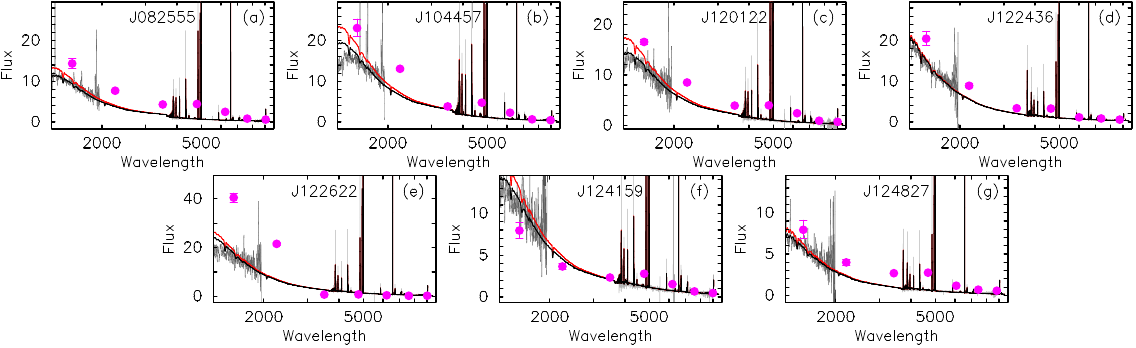}
\caption{The same as in Fig.~\ref{fig4} but for galaxies from \citet{Be16}. 
\label{figa2}}
\end{figure*}





\begin{figure*}
\includegraphics[angle=0,width=0.99\linewidth]{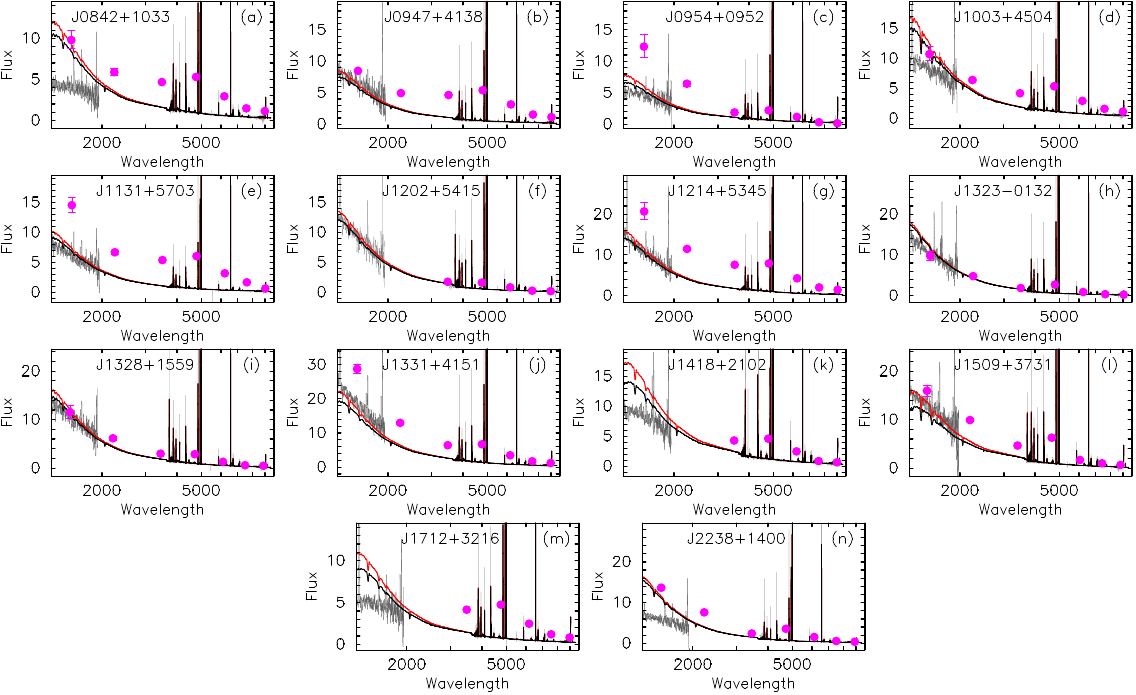}
\caption{The same as in Fig.~\ref{fig3} but for galaxies from \citet{Be19}. 
\label{figa3}}
\includegraphics[angle=0,width=0.99\linewidth]{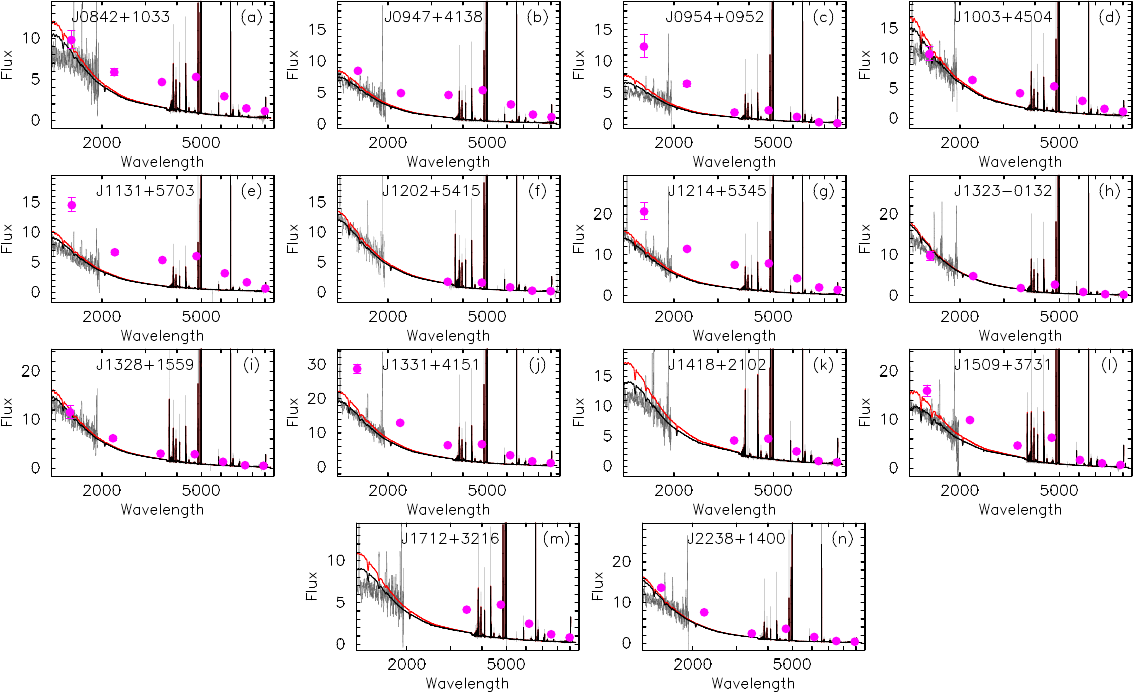}
\caption{The same as in Fig.~\ref{fig4} but for galaxies from \citet{Be19}. 
\label{figa4}}
\end{figure*}


\begin{figure*}
\includegraphics[angle=0,width=0.99\linewidth]{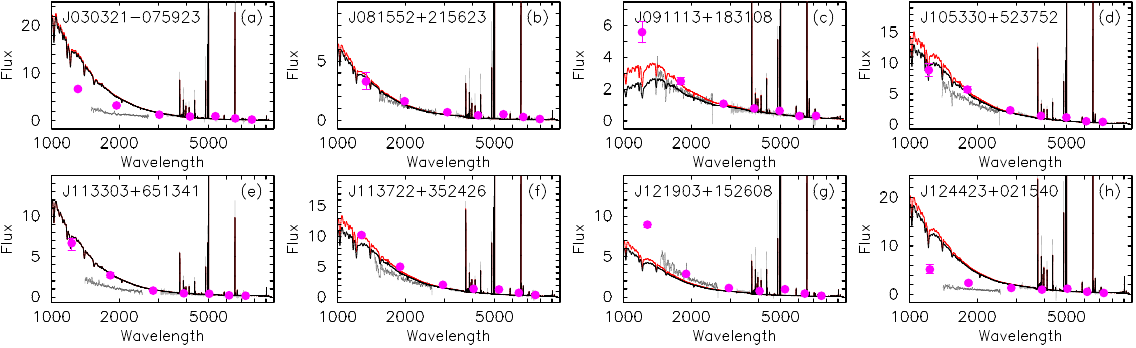}
\caption{The same as in Fig.~\ref{fig3} but for galaxies from \citet{Ra20}.
\label{figa5}}
\includegraphics[angle=0,width=0.99\linewidth]{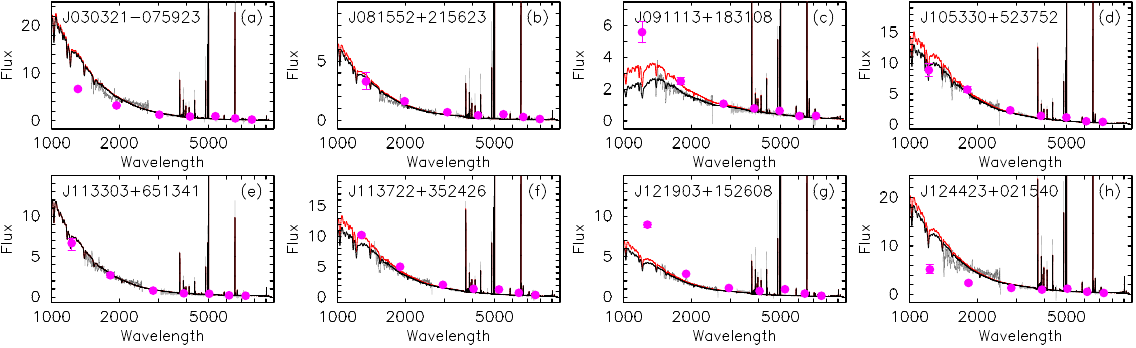}
\caption{The same as in Fig.~\ref{fig4} but for galaxies from \citet{Ra20}.
\label{figa6}}
\end{figure*}

\begin{figure*}
\includegraphics[angle=0,width=0.99\linewidth]{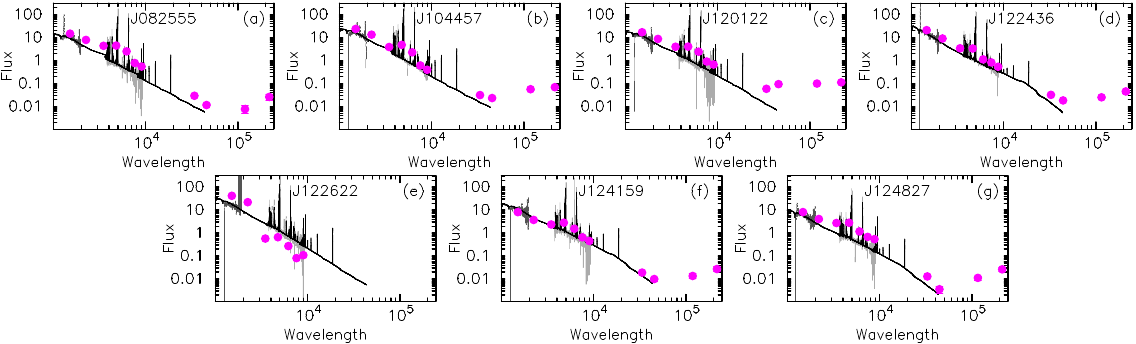}
\caption{The same as in Fig.~\ref{fig5} but for galaxies from \citet{Be16}.
Stellar and nebular SEDs are not shown.
\label{figa7}}
\end{figure*}

\begin{figure*}
\includegraphics[angle=0,width=0.99\linewidth]{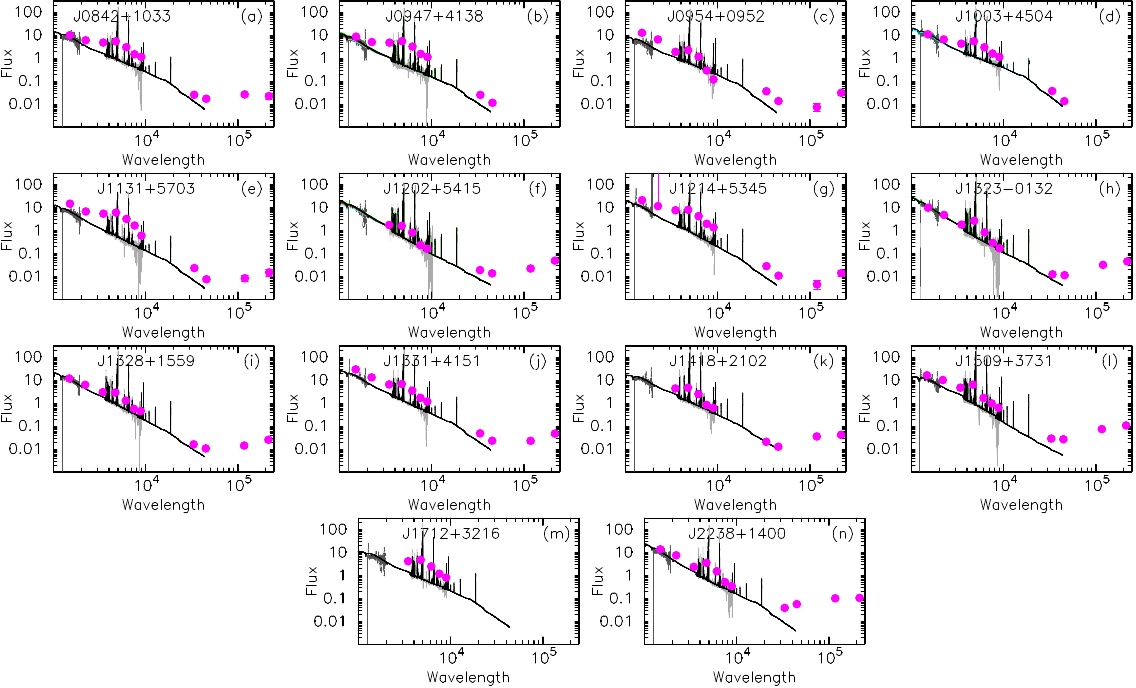}
\caption{The same as in Fig.~\ref{fig5} but for galaxies from \citet{Be19}.
Stellar and nebular SEDs are not shown.
\label{figa8}}
\end{figure*}

\begin{figure*}
\includegraphics[angle=0,width=0.99\linewidth]{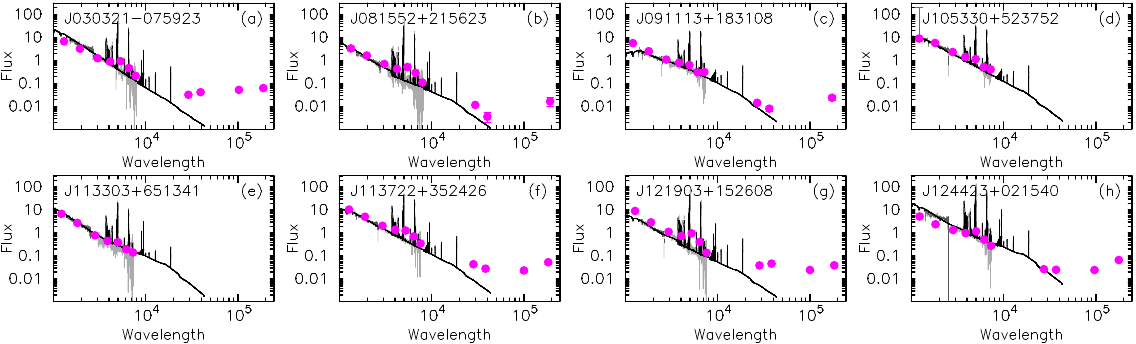}
\caption{The same as in Fig.~\ref{fig5} but for galaxies from \citet{Ra20}.
Stellar and nebular SEDs are not shown.
\label{figa9}}
\end{figure*}



\section{Characteristics for the determination of the C/O abundance ratios}

  \begin{table*}
  \caption{Characteristics for the determination of the C/O abundance ratio in galaxies from \citet{Be16} \label{taba1}}
\begin{tabular}{lcrcrrrcccccc} \hline
Name        &$F$$^{\rm a}$&EW$^{\rm b}$&$F$$^{\rm a}$&EW$^{\rm b}$&\multicolumn{1}{c}{$F$$^{\rm a}$}&EW$^{\rm b}$&$C_{\rm int}$$^{\rm c}$&$C_{\rm int}$$^{\rm c}$&$C_{\rm int}$$^{\rm c}$&corr.$^{\rm d}$ \\
            &(C~{\sc iii}]&(C~{\sc iii}]&(O~{\sc iii}]&(O~{\sc iii}]&\multicolumn{1}{c}{(H$\beta$}&\multicolumn{1}{c}{(H$\beta$}&(H$\beta$&(C~{\sc iii}]&(O~{\sc iii}]& \\
            &$\lambda$1908)&$\lambda$1908)&$\lambda$1666)&$\lambda$1666)&\multicolumn{1}{c}{$\lambda$4861)}&$\lambda$4861)&$\lambda$4861)&$\lambda$1908)&$\lambda$1666)&\\ \hline

J082555&133.4$\pm$16.0&16.7$\pm$2.0&35.3$\pm$16.0& 3.6$\pm$1.6&168.8$\pm$2.1&254$\pm$3&0.148$\pm$0.027&0.354$\pm$0.064&0.365$\pm$0.067&0.820\\
J104457&186.2$\pm$16.1&16.8$\pm$1.5&75.6$\pm$16.0& 6.3$\pm$1.3&307.0$\pm$2.7&285$\pm$2&0.176$\pm$0.027&0.421$\pm$0.065&0.437$\pm$0.067&1.000\\
J120122& 59.2$\pm$16.0&15.7$\pm$4.2&16.0$\pm$16.0& 3.0$\pm$3.0&188.2$\pm$1.1&201$\pm$1&0.176$\pm$0.027&0.421$\pm$0.065&0.437$\pm$0.067&1.667\\
J124159& 42.4$\pm$ 8.0&12.7$\pm$2.4&18.1$\pm$ 8.0& 5.0$\pm$2.2&145.4$\pm$1.0&130$\pm$1&0.150$\pm$0.028&0.359$\pm$0.067&0.370$\pm$0.069&2.222\\
J122622& 32.1$\pm$ 3.2&10.5$\pm$1.0&10.1$\pm$ 3.2& 3.2$\pm$1.0&181.6$\pm$1.0&154$\pm$1&0.084$\pm$0.027&0.201$\pm$0.065&0.207$\pm$0.067&4.000\\
J122436&111.8$\pm$ 4.2& 9.6$\pm$0.4&25.6$\pm$ 4.1& 1.8$\pm$0.3&108.3$\pm$1.9&116$\pm$2&0.024$\pm$0.028&0.057$\pm$0.068&0.059$\pm$0.069&0.741\\
J124827& 54.7$\pm$ 6.0&14.6$\pm$1.6&55.3$\pm$ 6.0& 2.4$\pm$0.3& 54.7$\pm$1.4&111$\pm$3&0.088$\pm$0.031&0.211$\pm$0.074&0.217$\pm$0.076&0.909\\
\hline
\end{tabular}

\begin{tabular}{lcccccccc} \hline
Name        &$t$(O~{\sc iii})$^{\rm e}$&\multicolumn{4}{c}{$I$($\lambda$)/$I$(H$\beta$)$^{\rm f}$}&12+log(O/H)&log(C/O)$_{\rm opt}$$^{\rm g}$&log(C/O)$_{\rm UV}$$^{\rm h}$\\
            &                &(C~{\sc iii}]&(O~{\sc iii}]&([O~{\sc ii}] &([O~{\sc iii}]&\\ 
            &                &$\lambda$1908)&$\lambda$1666)&$\lambda$3727)&$\lambda$5007)&\\ \hline
J082555&1.89$\pm$0.06&1.042$\pm$0.125&0.283$\pm$0.128&0.309$\pm$0.014&3.614$\pm$0.075&7.41$\pm$0.01& $-$0.584$\pm$0.060&$-$0.470$\pm$0.085\\
J104457&1.96$\pm$0.05&1.066$\pm$0.052&0.446$\pm$0.052&0.331$\pm$0.011&4.336$\pm$0.089&7.46$\pm$0.01& $-$0.684$\pm$0.060&$-$0.640$\pm$0.085\\
J120122&1.76$\pm$0.05&0.922$\pm$0.249&0.257$\pm$0.069&0.539$\pm$0.012&3.559$\pm$0.075&7.50$\pm$0.01& $-$0.594$\pm$0.110&$-$0.542$\pm$0.155\\
J124159&1.64$\pm$0.05&1.048$\pm$0.198&0.459$\pm$0.203&1.182$\pm$0.008&4.610$\pm$0.098&7.68$\pm$0.01& $-$0.605$\pm$0.160&$-$0.796$\pm$0.226\\
J122622&1.63$\pm$0.04&0.926$\pm$0.277&0.295$\pm$0.093&0.929$\pm$0.020&5.576$\pm$0.116&7.77$\pm$0.01& $-$0.702$\pm$0.050&$-$0.628$\pm$0.071\\
J122436&1.55$\pm$0.05&0.821$\pm$0.031&0.192$\pm$0.032&0.821$\pm$0.024&5.550$\pm$0.120&7.82$\pm$0.01& $-$0.680$\pm$0.070&$-$0.501$\pm$0.099\\
J124827&1.59$\pm$0.07&1.219$\pm$0.133&0.252$\pm$0.027&0.733$\pm$0.026&6.049$\pm$0.133&7.82$\pm$0.01& $-$0.564$\pm$0.170&$-$0.425$\pm$0.240\\
\hline
\end{tabular}

\hbox{$^{\rm a}$Observed flux in 10$^{-16}$ erg s$^{-1}$ cm$^{-2}$.}

\hbox{$^{\rm b}$Restframe equivalent width in \AA.}

\hbox{$^{\rm c}$Extinction coefficient for H$\beta$, C~{\sc iii}] $\lambda$1908
and O~{\sc iii}] $\lambda$1666 emission lines.}

\hbox{$^{\rm d}$Correction factor used to adjust observed UV spectrum with the modelled SED.}

  \hbox{$^{\rm e}$$t$(O~{\sc iii}) = $T_{\rm e}$(O~{\sc iii})/10$^4$K, where $T_{\rm e}$(O~{\sc iii}) is the electron temperature in O$^{2+}$ zone derived by the
    direct method.}

\hbox{$^{\rm f}$Extinction-corrected flux ratios.}

\hbox{$^{\rm g}$C/O abundance ratio is derived using C~{\sc iii}] $\lambda$1908 and [O~{\sc iii}] $\lambda$4959,5007 emission-line fluxes.}

\hbox{$^{\rm h}$C/O abundance ratio is derived using C~{\sc iii}] $\lambda$1908 and O~{\sc iii}] $\lambda$1666 emission-line fluxes.}

  \end{table*}

  \begin{table*}
  \caption{Characteristics for the determination of the C/O abundance ratio in galaxies from \citet{Be19} \label{taba2}}
\begin{tabular}{lcrcrrrcccccc} \hline
Name        &$F$$^{\rm a}$&EW$^{\rm b}$&$F$$^{\rm a}$&EW$^{\rm b}$&\multicolumn{1}{c}{$F$$^{\rm a}$}&EW$^{\rm b}$&$C_{\rm int}$$^{\rm c}$&$C_{\rm int}$$^{\rm c}$&$C_{\rm int}$$^{\rm c}$&corr.$^{\rm d}$ \\
            &(C~{\sc iii}]&(C~{\sc iii}]&(O~{\sc iii}]&(O~{\sc iii}]&\multicolumn{1}{c}{(H$\beta$}&\multicolumn{1}{c}{(H$\beta$}&(H$\beta$&(C~{\sc iii}]&(O~{\sc iii}]& \\
            &$\lambda$1908)&$\lambda$1908)&$\lambda$1666)&$\lambda$1666)&\multicolumn{1}{c}{$\lambda$4861)}&$\lambda$4861)&$\lambda$4861)&$\lambda$1908)&$\lambda$1666)&\\ \hline

J084236& 28.9$\pm$12.0&13.3$\pm$5.5&20.1$\pm$12.0& 6.1$\pm$6.1&106.2$\pm$1.8&150$\pm$3&0.128$\pm$0.028&0.306$\pm$0.067&0.316$\pm$0.069&1.812\\
J094718& 80.2$\pm$16.0&22.0$\pm$4.4&14.8$\pm$16.0& 2.7$\pm$2.9& 89.6$\pm$1.5&219$\pm$4&0.116$\pm$0.029&0.278$\pm$0.070&0.286$\pm$0.072&0.897\\
J095430& 51.9$\pm$ 8.0&23.6$\pm$3.6&23.8$\pm$ 8.0& 6.7$\pm$2.3& 83.9$\pm$1.6&153$\pm$3&0.150$\pm$0.029&0.359$\pm$0.069&0.370$\pm$0.072&1.000\\
J100348& 66.9$\pm$ 8.0&14.7$\pm$1.8&23.7$\pm$ 8.0& 3.7$\pm$1.2& 98.2$\pm$1.8&100$\pm$2&0.108$\pm$0.029&0.258$\pm$0.069&0.266$\pm$0.074&1.177\\
J113116& 49.5$\pm$ 6.4&10.0$\pm$1.3&14.1$\pm$ 6.4& 2.8$\pm$1.3& 61.0$\pm$1.4&124$\pm$3&0.092$\pm$0.030&0.220$\pm$0.072&0.227$\pm$0.074&1.000\\
J120202& 69.1$\pm$ 7.2&12.4$\pm$1.3&27.5$\pm$ 7.2& 3.9$\pm$1.0&140.2$\pm$2.0&292$\pm$4&0.100$\pm$0.028&0.239$\pm$0.067&0.247$\pm$0.069&1.000\\
J121402&107.7$\pm$ 6.5&16.6$\pm$1.0&34.1$\pm$ 6.4& 3.9$\pm$0.7&119.3$\pm$1.9&130$\pm$2&0.112$\pm$0.028&0.268$\pm$0.067&0.276$\pm$0.069&1.000\\
J132347& 88.0$\pm$ 8.8&17.1$\pm$1.7&76.5$\pm$ 8.8&10.2$\pm$1.2&137.9$\pm$1.0&256$\pm$2&0.032$\pm$0.028&0.077$\pm$0.067&0.079$\pm$0.069&1.000\\
J132853& 50.5$\pm$12.8& 6.7$\pm$1.7&25.3$\pm$12.8& 3.1$\pm$1.6&132.5$\pm$2.0&171$\pm$3&0.084$\pm$0.032&0.201$\pm$0.077&0.207$\pm$0.079&1.000\\
J133126&163.5$\pm$16.1&15.7$\pm$1.5&80.5$\pm$16.0& 5.8$\pm$1.2&182.9$\pm$2.3&158$\pm$2&0.128$\pm$0.027&0.306$\pm$0.065&0.316$\pm$0.067&0.800\\
J141851&126.7$\pm$ 6.9&24.5$\pm$1.4&70.0$\pm$ 6.8&10.9$\pm$1.1&244.2$\pm$2.7&228$\pm$3&0.204$\pm$0.027&0.488$\pm$0.065&0.503$\pm$0.067&1.250\\
J150934&121.1$\pm$12.1&15.3$\pm$1.5&54.9$\pm$12.0& 5.6$\pm$1.2&190.1$\pm$2.4&237$\pm$3&0.220$\pm$0.027&0.526$\pm$0.067&0.542$\pm$0.067&0.833\\
J171236& 55.1$\pm$ 8.0&18.6$\pm$2.7&27.2$\pm$ 8.0& 7.1$\pm$2.1&108.6$\pm$1.7&179$\pm$3&0.175$\pm$0.031&0.419$\pm$0.074&0.431$\pm$0.076&1.333\\
J223831& 48.5$\pm$ 8.2&19.9$\pm$3.3&21.8$\pm$ 8.1& 4.2$\pm$1.5& 87.6$\pm$1.6&185$\pm$4&0.052$\pm$0.030&0.124$\pm$0.072&0.128$\pm$0.074&1.667\\
\hline
\end{tabular}

\begin{tabular}{lcccccccc} \hline
Name        &$t$(O~{\sc iii})$^{\rm e}$&\multicolumn{4}{c}{$I$($\lambda$)/$I$(H$\beta$)$^{\rm f}$}&12+log(O/H)&log(C/O)$_{\rm opt}$$^{\rm g}$&log(C/O)$_{\rm UV}$$^{\rm h}$\\
            &                &(C~{\sc iii}]&(O~{\sc iii}]&([O~{\sc ii}] &([O~{\sc iii}]&\\ 
            &                &$\lambda$1908)&$\lambda$1666)&$\lambda$3727)&$\lambda$5007)&\\ \hline
J084236&1.79$\pm$0.06&0.743$\pm$0.309&0.529$\pm$0.316&0.426$\pm$0.016&5.150$\pm$0.110&7.63$\pm$0.01& $-$0.823$\pm$0.114&$-$0.900$\pm$0.161\\
J094718&1.61$\pm$0.05&1.166$\pm$0.233&0.219$\pm$0.237&0.698$\pm$0.020&5.411$\pm$0.116&7,76$\pm$0.01& $-$0.556$\pm$0.079&$-$0.384$\pm$0.111\\
J095430&1.69$\pm$0.06&1.001$\pm$0.154&0.471$\pm$0.078&0.590$\pm$0.019&5.418$\pm$0.116&7.71$\pm$0.01& $-$0.669$\pm$0.111&$-$0.757$\pm$0.157\\
J100348&1.57$\pm$0.05&1.133$\pm$0.136&0.409$\pm$0.050&0.452$\pm$0.017&6.831$\pm$0.143&7.87$\pm$0.01& $-$0.587$\pm$0.129&$-$0.626$\pm$0.182\\
J113116&1.67$\pm$0.07&1.089$\pm$0.141&0.316$\pm$0.143&0.872$\pm$0.035&5.678$\pm$0.126&7.72$\pm$0.01& $-$0.664$\pm$0.142&$-$0.575$\pm$0.200\\
J120202&1.83$\pm$0.06&0.679$\pm$0.071&0.275$\pm$0.057&0.529$\pm$0.016&3.660$\pm$0.078&7.48$\pm$0.01& $-$0.780$\pm$0.093&$-$0.690$\pm$0.131\\
J121402&1.74$\pm$0.06&1.293$\pm$0.078&0.417$\pm$0.078&0.646$\pm$0.018&5.137$\pm$0.108&7.65$\pm$0.01& $-$0.580$\pm$0.102&$-$0.595$\pm$0.144\\
J132347&1.74$\pm$0.04&0.708$\pm$0.071&0.619$\pm$0.071&0.218$\pm$0.012&7.222$\pm$0.152&7.78$\pm$0.01& $-$0.873$\pm$0.079&$-$0.912$\pm$0.111\\
J132853&1.60$\pm$0.05&0.475$\pm$0.063&0.241$\pm$0.122&0.829$\pm$0.023&5.002$\pm$0.105&7.75$\pm$0.01& $-$0.921$\pm$0.116&$-$0.836$\pm$0.164\\
J133126&1.63$\pm$0.04&1.077$\pm$0.103&0.542$\pm$0.108&0.701$\pm$0.019&5.566$\pm$0.115&7.76$\pm$0.01& $-$0.616$\pm$0.075&$-$0.807$\pm$0.106\\
J141851&1.86$\pm$0.05&1.247$\pm$0.068&0.713$\pm$0.069&0.410$\pm$0.013&4.634$\pm$0.095&7.56$\pm$0.01& $-$0.600$\pm$0.065&$-$0.800$\pm$0.092\\
J150934&1.61$\pm$0.04&0.932$\pm$0.093&0.505$\pm$0.110&0.483$\pm$0.014&6.869$\pm$0.143&7.84$\pm$0.01& $-$0.711$\pm$0.087&$-$0.799$\pm$0.123\\
J171236&1.52$\pm$0.05&1.186$\pm$0.172&0.361$\pm$0.106&0.738$\pm$0.020&5.911$\pm$0.126&7.86$\pm$0.01& $-$0.510$\pm$0.102&$-$0.609$\pm$0.144\\
J223831&1.80$\pm$0.06&1.090$\pm$0.185&0.494$\pm$0.184&0.307$\pm$0.014&4.902$\pm$0.094&7.60$\pm$0.01& $-$0.620$\pm$0.085&$-$0.681$\pm$0.120\\
\hline
\end{tabular}

\hbox{$^{\rm a}$Observed flux in 10$^{-16}$ erg s$^{-1}$ cm$^{-2}$.}

\hbox{$^{\rm b}$Restframe equivalent width in \AA.}

\hbox{$^{\rm c}$Extinction coefficient for H$\beta$, C~{\sc iii}] $\lambda$1908
and O~{\sc iii}] $\lambda$1666 emission lines.}

\hbox{$^{\rm d}$Correction factor used to adjust observed UV spectrum with the modelled SED.}

  \hbox{$^{\rm e}$$t$(O~{\sc iii}) = $T_{\rm e}$(O~{\sc iii})/10$^4$K, where $T_{\rm e}$(O~{\sc iii}) is the electron temperature in O$^{2+}$ zone derived by the
    direct method.}

\hbox{$^{\rm f}$Extinction-corrected flux ratios.}

\hbox{$^{\rm g}$C/O abundance ratio is derived using C~{\sc iii}] $\lambda$1908 and [O~{\sc iii}] $\lambda$4959,5007 emission-line fluxes.}

\hbox{$^{\rm h}$C/O abundance ratio is derived using C~{\sc iii}] $\lambda$1908 and O~{\sc iii}] $\lambda$1666 emission-line fluxes.}

  \end{table*}

  \begin{table*}
  \caption{Characteristics for the determination of the C/O abundance ratio in galaxies from \citet{Ra20} \label{taba3}}
\begin{tabular}{lcrcrrrcccccc} \hline
Name        &$F$$^{\rm a}$&EW$^{\rm b}$&$F$$^{\rm a}$&EW$^{\rm b}$&\multicolumn{1}{c}{$F$$^{\rm a}$}&EW$^{\rm b}$&$C_{\rm int}$$^{\rm c}$&$C_{\rm int}$$^{\rm c}$&$C_{\rm int}$$^{\rm c}$&corr.$^{\rm d}$ \\
            &(C~{\sc iii}]&(C~{\sc iii}]&(O~{\sc iii}]&(O~{\sc iii}]&\multicolumn{1}{c}{(H$\beta$}&\multicolumn{1}{c}{(H$\beta$}&(H$\beta$&(C~{\sc iii}]&(O~{\sc iii}]& \\
            &$\lambda$1908)&$\lambda$1908)&$\lambda$1666)&$\lambda$1666)&\multicolumn{1}{c}{$\lambda$4861)}&$\lambda$4861)&$\lambda$4861)&$\lambda$1908)&$\lambda$1666)&\\ \hline

J030321-075923& 10.3$\pm$1.1& 8.0$\pm$0.9&...&...& 62.7$\pm$0.8&140$\pm$2&0.024$\pm$0.030&0.078$\pm$0.098&...&3.333\\
J081552+215623& 13.0$\pm$1.6&10.4$\pm$1.3&...&...& 34.3$\pm$1.1&257$\pm$8&0.060$\pm$0.033&0.144$\pm$0.079&...&1.000\\
J091113+183108&  4.9$\pm$1.1& 2.5$\pm$0.6&...&...& 37.7$\pm$1.3& 83$\pm$3&0.336$\pm$0.031&0.804$\pm$0.074&...&0.934\\
J105330+523752&  6.1$\pm$2.3& 1.9$\pm$0.7&...&...& 51.4$\pm$1.5& 92$\pm$2&0.120$\pm$0.031&0.287$\pm$0.074&...&1.333\\
J113303+651341&  4.2$\pm$1.1& 4.2$\pm$1.1&...&...& 24.1$\pm$0.5&110$\pm$2&0.000&0.000&...&2.326\\
J113722+352426&  9.7$\pm$2.5& 3.3$\pm$0.8&...&...& 65.8$\pm$1.6&128$\pm$3&0.124$\pm$0.030&0.297$\pm$0.072&...&1.250\\
J121903+152608& 30.2$\pm$2.3&13.3$\pm$1.0&...&...& 62.1$\pm$1.5&279$\pm$7&0.048$\pm$0.030&0.115$\pm$0.072&...&0.714\\
J124423+021540&  8.5$\pm$1.1& 8.7$\pm$1.2&...&...&126.0$\pm$2.1&197$\pm$3&0.072$\pm$0.028&0.172$\pm$0.067&...&5.000\\
J124834+123402&  7.9$\pm$1.2& 5.9$\pm$0.9&...&...& 21.1$\pm$0.8&144$\pm$6&0.065$\pm$0.064&0.156$\pm$0.154&...&1.176\\
J145735+223201& 18.3$\pm$1.2&16.3$\pm$1.1&...&...& 90.0$\pm$1.8&235$\pm$5&0.160$\pm$0.029&0.383$\pm$0.069&...&2.222\\
\hline
\end{tabular}

\begin{tabular}{lcccccccc} \hline
Name        &$t$(O~{\sc iii})$^{\rm e}$&\multicolumn{4}{c}{$I$($\lambda$)/$I$(H$\beta$)$^{\rm f}$}&12+log(O/H)&log(C/O)$_{\rm opt}$$^{\rm g}$&log(C/O)$_{\rm UV}$$^{\rm h}$\\
            &                &(C~{\sc iii}]&(O~{\sc iii}]&([O~{\sc ii}] &([O~{\sc iii}]&\\ 
            &                &$\lambda$1908)&$\lambda$1666)&$\lambda$3727)&$\lambda$5007)&\\ \hline
J030321-075923&1.53$\pm$0.06&0.611$\pm$0.065&...&0.986$\pm$0.031&5.816$\pm$0.127&7.86$\pm$0.01& $-$0.821$\pm$0.112&...\\
J081552+215623&1.43$\pm$0.06&0.459$\pm$0.057&...&0.701$\pm$0.029&7.118$\pm$0.160&8.00$\pm$0.01& $-$0.902$\pm$0.112&...\\
J091113+183108&1.26$\pm$0.13&0.357$\pm$0.080&...&2.006$\pm$0.063&3.519$\pm$0.086&7.99$\pm$0.03& $-$0.650$\pm$0.231&...\\
J105330+523752&1.22$\pm$0.09&0.232$\pm$0.052&...&1.973$\pm$0.056&4.570$\pm$0.104&8.10$\pm$0.02& $-$0.872$\pm$0.229&...\\
J113303+651341&1.41$\pm$0.10&0.407$\pm$0.106&...&1.460$\pm$0.054&5.751$\pm$0.141&7.96$\pm$0.02& $-$0.910$\pm$0.152&...\\
J113722+352426&1.19$\pm$0.06&0.274$\pm$0.071&...&1.784$\pm$0.049&5.025$\pm$0.112&8.16$\pm$0.02& $-$0.783$\pm$0.106&...\\
J121903+152608&1.51$\pm$0.05&0.405$\pm$0.031&...&0.628$\pm$0.022&6.417$\pm$0.140&7.89$\pm$0.01& $-$0.986$\pm$0.096&...\\
J124423+021540&1.21$\pm$0.04&0.425$\pm$0.057&...&1.476$\pm$0.038&5.845$\pm$0.125&8.17$\pm$0.01& $-$0.660$\pm$0.129&...\\
J124834+123402&1.26$\pm$0.10&0.543$\pm$0.082&...&1.606$\pm$0.094&6.888$\pm$0.287&8.12$\pm$0.08& $-$0.629$\pm$0.138&...\\
J145735+223201&1.41$\pm$0.04&0.755$\pm$0.049&...&0.970$\pm$0.028&7.333$\pm$0.154&8.04$\pm$0.01& $-$0.700$\pm$0.100&...\\
\hline
\end{tabular}

\hbox{$^{\rm a}$Observed flux in 10$^{-16}$ erg s$^{-1}$ cm$^{-2}$.}

\hbox{$^{\rm b}$Restframe equivalent width in \AA.}

\hbox{$^{\rm c}$Extinction coefficient for H$\beta$, C~{\sc iii}] $\lambda$1908
and O~{\sc iii}] $\lambda$1666 emission lines.}

\hbox{$^{\rm d}$Correction factor used to adjust observed UV spectrum with the modelled SED.}

  \hbox{$^{\rm e}$$t$(O~{\sc iii}) = $T_{\rm e}$(O~{\sc iii})/10$^4$K, where $T_{\rm e}$(O~{\sc iii}) is the electron temperature in O$^{2+}$ zone derived by the
    direct method.}

\hbox{$^{\rm f}$Extinction-corrected flux ratios.}

\hbox{$^{\rm g}$C/O abundance ratio is derived using C~{\sc iii}] $\lambda$1908 and [O~{\sc iii}] $\lambda$4959,5007 emission-line fluxes.}

\hbox{$^{\rm h}$C/O abundance ratio is derived using C~{\sc iii}] $\lambda$1908 and O~{\sc iii}] $\lambda$1666 emission-line fluxes.}

  \end{table*}

\bsp	
\label{lastpage}
\end{document}